\documentclass[twocolumn]{aastex63}

\shorttitle{Identification of single spectral lines using \umlaut, a new software}
\shortauthors{Baronchelli I.}

\accepted{for publication on ApJS, 2021 September 6}

\usepackage{graphicx}
\usepackage{CJK} 

\newcommand{\mic}{$\mu$m}
\newcommand{\ha}{H$\alpha$}

\newcommand{\oiii}{[O\thinspace{\sc iii}]}
\newcommand{\oii}{[O\thinspace{\sc ii}]}

\newcommand{\umlaut}{\footnotesize{\emph{UMLAUT}}\normalsize}

\DeclareGraphicsExtensions{.pdf,.png,.jpg}

\begin{document}
\begin{CJK*}{UTF8}{gbsn} 

 \title{Identification of single spectral lines in large spectroscopic surveys using \umlaut: an Unsupervised Machine Learning Algorithm based on Unbiased Topology}

\correspondingauthor{I. Baronchelli}
\email{ivano.baronchelli@inaf.it}

 \author{I. Baronchelli}
\affiliation{$INAF-$Osservatorio Astronomico di Padova, Vicolo dell'Osservatorio 5, I-35122, Padova, Italy}
\affiliation{Dipartimento di Fisica e Astronomia, Universit${\grave{\mathrm{a}}}$ di Padova, vicolo Osservatorio, 3, 35122 Padova, Italy.}

\author{C. M. Scarlata}
\affiliation{MN Institute for Astrophysics, University of Minnesota, 116 Church Street SE,  Minneapolis, MN 55455, USA.}

\author{L. Rodr\'{\i}guez-Mu\~noz}
\affiliation{Dipartimento di Fisica e Astronomia, Universit${\grave{\mathrm{a}}}$ di Padova, vicolo Osservatorio, 3, 35122 Padova, Italy.}

\author{M. Bonato}
\affiliation{$INAF-$Istituto di Radioastronomia and Italian ALMA Regional Centre, Via Gobetti 101, I-40129, Bologna, Italy}
 \affiliation{$INAF-$Osservatorio Astronomico di Padova, Vicolo dell'Osservatorio 5, I-35122, Padova, Italy}

\author{L. Morselli}
\affiliation{$INAF-$Osservatorio Astronomico di Padova, Vicolo dell'Osservatorio 5, I-35122, Padova, Italy}
\affiliation{Dipartimento di Fisica e Astronomia, Universit${\grave{\mathrm{a}}}$ di Padova, vicolo Osservatorio, 3, 35122 Padova, Italy.}

\author{M. Vaccari}
\affiliation{Inter-University Institute for Data Intensive Astronomy (IDIA) - Department of Physics \& Astronomy, South Africa}
\affiliation{University of the Western Cape, Robert Sobukwe Road, 7535 Bellville, Cape Town, South Africa}

\author{R. Carraro}
\affiliation{Instituto de F\'\i{}sica y Astronom\'\i{}a, Universidad de Valpara\'\i{}so, Gran Breta\~{n}a 1111, Playa Ancha, Valpara\'\i{}so, Chile}

\author{L. Barrufet}
\affiliation{Universit\'e de Gen\`eve, Department of Astronomy, Chemin Pegasi, 51, 1290 Versoix, Switzerland}

\author{A. Henry}
\affiliation{Space Telescope Science Institute, 3700 San Martin Drive, Baltimore, MD, 21218, USA}

\author{V. Mehta}
\affiliation{MN Institute for Astrophysics, University of Minnesota, 116 Church St. SE,  Minneapolis, MN 55455, USA.}

\author{G. Rodighiero}
\affiliation{Dipartimento di Fisica e Astronomia, Universit${\grave{\mathrm{a}}}$ di Padova, vicolo Osservatorio, 3, 35122 Padova, Italy.}

\author{A. Baruffolo}
\affiliation{$INAF-$Osservatorio Astronomico di Padova, Vicolo dell'Osservatorio 5, I-35122, Padova, Italy}

\author{M. Bagley}
\affiliation{College of Natural Sciences, The University of Texas at Austin, 2515 Speedway, Austin, TX 78712, USA}

\author{A. Battisti}
\affiliation{Research School of Astronomy and Astrophysics, Australian National University, Cotter Road, Weston Creek, ACT 2611, Australia}

\author{J. Colbert}
\affiliation{IPAC, Mail Code 314-6, Caltech, 1200 E. California Blvd., Pasadena, CA 91125, USA.}

\author{Y. S. Dai(戴昱)}
\affiliation{Chinese Academy of Sciences South America Center for Astronomy (CASSACA)/NAOC, 20A Datun Road, Beijing 100101, China}

\author{M. De Pascale}
\affiliation{$INAF-$Osservatorio Astronomico di Padova, Vicolo dell'Osservatorio 5, I-35122, Padova, Italy}

\author{H. Dickinson}
\affiliation{School of Physical Sciences, The Open University, Walton Hall, Milton Keynes, MK7 6AA, UK}

\author{M. Malkan}
\affiliation{Department of Physics and Astronomy, UCLA, Physics and Astronomy Bldg., 3-714, LA CA 90095-1547, USA}

\author{C. Mancini}
\affiliation{Dipartimento di Fisica e Astronomia, Universit${\grave{\mathrm{a}}}$ di Padova, vicolo Osservatorio, 3, 35122 Padova, Italy.}
\affiliation{INAF-IASF, Via Alfonso Corti 12 I-20133 Milano Italy}

\author{M. Rafelski}
\affiliation{Space Telescope Science Institute, 3700 San Martin Drive, Baltimore, MD 21218, USA}
\affiliation{Department of Physics and Astronomy, Johns Hopkins University, Baltimore, MD 21218, USA}

\author{H. I. Teplitz}
\affiliation{IPAC, Mail Code 314-6, Caltech, 1200 E. California Boulevard, Pasadena, CA 91125, USA}

\begin{abstract}

{\small The identification of an emission line is unambiguous when multiple spectral features are clearly visible in the same spectrum.}{\small  However, in many cases, only one line is detected, making it difficult to correctly determine the redshift.}{\small  We developed a freely available unsupervised machine-learning algorithm based on unbiased topology (\umlaut) that can be used in a very wide variety of contexts, including the identification of single emission lines.}{\small   To this purpose, the algorithm combines different sources of information, such as the apparent magnitude, size and color of the emitting source, and the equivalent width and wavelength of the detected line.}{\small  In each specific case, the algorithm automatically identifies the most relevant ones (i.e., those able to minimize the dispersion associated with the output parameter).}{\small  The outputs can be easily integrated into different algorithms, allowing us to combine supervised and unsupervised techniques and increasing the overall accuracy.}{\small  We tested our software on WISP (WFC3 IR Spectroscopic Parallel) survey data.}{\small  WISP represents one of the closest existing analogs to the near-IR spectroscopic surveys that are going to be performed by the future Euclid and Roman missions.}{\small  These missions will investigate the large-scale structure of the universe by surveying a large portion of the extragalactic sky in near-IR slitless spectroscopy, detecting a relevant fraction of single emission lines.}{\small  In our tests, \umlaut\  correctly identifies real lines in 83.2\% of the cases.}{\small  The accuracy is slightly higher (84.4\%) when combining our unsupervised approach with a supervised approach we previously developed.} \\ \\

\end{abstract}


\section{Introduction}
\label{introduction}

Determining the spectroscopic redshift of a source represents an easy task when some general conditions are satisfied. In particular, spectral lines are securely identified when i) the signal-to-noise ratio (S/N) is high, ii) there is no (or negligible) contamination from close sources, iii) multiple spectral features are clearly visible in the same spectrum. Satisfying all these conditions becomes increasingly difficult when fainter and fainter sources are considered and/or when the wavelength range studied is particularly narrow. 

With the goal of probing the nature of dark energy, the future Euclid and Roman missions are going to obtain a huge amount of near-IR slitless spectra over an unprecedented area in the sky \citep{2011arXiv1110.3193L,2012SPIE.8442E..0TL,2012arXiv1208.4012G,2015arXiv150303757S}. In this context, the automatic identification of single emission lines is going to be crucial.

In a recent paper \citet[][BA20 hereafter]{2020ApJS..249...12B}, we described a \emph{supervised} machine-learning (ML) algorithm that we used to identify single lines ($\sim82.6$\% accuracy on real lines) in a Hubble Space Telescope (HST) near-IR spectroscopic survey; \citep[WISPs, WFC3 Infra-red Spectroscopic Parallel survey][]{2010ApJ...723..104A}. WISPs represents one of the most important proxies for future Euclid and Roman surveys, both in terms of spectrophotometric coverage and for the particularly wide sky area covered ($\sim$1520 arcmin$^{2}$).

The algorithm of BA20 is organized in modules and blocks so that it can be easily adapted to include new ancillary information and/or different strategies. However, the purpose of such an algorithm is not meant to be general, and the software is designed for a pretty narrow category of problems (specifically, the identification of emission lines in spectroscopic surveys). In fact, following the \emph{supervised} paradigm, the algorithm implements preset functions relating the input parameters (apparent magnitude, size, color indexes, etc...) with the outputs (the nature of the spectral lines detected and the redshift of the emitting sources). In this case, what the algorithm \emph{learns} from the data are the parameters defining these functions.

In this paper, we describe and test an alternative \emph{unsupervised} machine-learning algorithm based on unbiased topology\footnote{ In the context of data collection, a sample is ``biased'' when the probability to select one or more subsets is systematically higher or lower than what is expected from a random selection. In an alternative definition, an estimator is said to be ``biased'' when it systematically underfits the data of the test set \citep[see, e.g. Section 8.11 in ][]{2014sdmm.book.....I}, for example, when a linear fit is used to describe a more complex distribution of data. \umlaut\ performs a topological analysis on the input data using a variant of the K-NN algorithm \citep[see, e.g.][]{doi:10.1080/00031305.1992.10475879}. Such an estimator is ``unbiased'' but subject to high variance when the parameter space is undersampled \citep[see, e.g. Section 9.4 in ][]{2014sdmm.book.....I} } (\umlaut) that we developed. Under this different paradigm, the algorithm automatically learns \emph{if} and \emph{how} the input information is structured and how it relates (or not) to the outputs. In this context, no preset functions are provided by the user.

We anticipate that the \emph{unsupervised} algorithm presented in this paper provides accuracies similar to those reached in BA20. However, differently from the algorithm of BA20, \umlaut\  is highly flexible and applicable to a very wide variety of problems.
Furthermore, the outputs of \umlaut\  can be combined with those of BA20, providing higher accuracies than when the two algorithms are considered separately.
 \umlaut\  is freely available at https://github.com/Ibaronch/UMLAUT together with additional documentation and examples of use.

We highlight the fact that the ML algorithms discussed in this paper are not simply used to estimate photometric redshifts. Under this point of view, in \cite{2020A&A...644A..31E} the reader can find the results of a recent challenge performed on a Euclid-like dataset and involving various ML algorithms. Instead, differently from a simpler photo-$z$ technique, in this paper we combine photometric and spectroscopic analyses. Moreover, the output of the algorithm is not a photo-$z$ estimation, but the identification of a single spectral line (with an associated probability).

The paper is organized as follows: in Section~\ref{SEC:DATA} we present the WISP survey. The characteristics of the training and test sets are also discussed. In Section~\ref{SECT:THE_ALGORTHM} we describe \umlaut, its features, and its internal behavior. In Section~\ref{SECT:WISP_TESTS} we compare the performances of \umlaut\  and of the BA20 supervised algorithm, using the same WISP data sample as a test bench. In Section~\ref{SECT:DISCUSSION} we summarize the advantages and limitations of \umlaut, also in comparison with different techniques. We draw the conclusions in Section~\ref{SECT:SUMMARY_AND_CONL}.

\section{Data}
\label{SEC:DATA}
In order to compare our unsupervised ML approach with the supervised strategy of BA20, we tested \umlaut\  on the same data sample, i.e., on the second data release of the WISP survey \citep{2010ApJ...723..104A,2020ApJ...897...98B,2020ApJS..249...12B}\footnote{https://archive.stsci.edu/prepds/wisp/}. In this section, we briefly outline the properties of the data sample used for our tests. A more extensive discussion can be found in BA20.

\subsection{The WISP survey}
\label{SEC:WISP}

WISP is a purely parallel HST/WFC3 slitless survey that collects near-infrared grism spectra in the wavelength ranges 0.8-1.1 \mic\ (G102, $R\sim$210) and 1.07-1.7 \mic\ (G141, $R\sim$130), approximately corresponding to the $J$ and $H$ photometric bands respectively. The median 5$\sigma$ depth reached in both grisms is $5\times 10^{-17}$ erg s$^{-1}$ cm$^{-2}$, with a factor of approximately 2 field-to-field variation.

For a large fraction of the fields, direct images in the $J$ and $H$ bands (F110W and F140W/F160W) are collected as well, and observations in the WFC3/UVIS bands and in the 3.6 \mic\ and 4.5 \mic\ channels of IRAC/\emph{Spitzer} are available (see Figure 3 in BA20 for precise statistics). Thanks to its spectral coverage, WISP can detect \ha\ from $z\sim 0.3$ to $z\sim 1.5$ and \oiii\ from $z\sim 0.7$ to $z\sim2.3$ (see Figure 23 in BA20).

In this work, we use data collected in 419 WISP fields, collectively covering a total area of 1520 arcmin$^{2}$.
Being a purely parallel survey, WISP exposures are taken while HST is engaged in observations of different primary targets with instruments other than WFC3. This approach guarantees a random selection of the WISP fields and strongly limits the possibility of spatially based biases.

In order to reduce the number of spurious identifications, the line detection is performed through a wavelet convolution and a SExtractor-type threshold, using a custom line-finding software. Three contiguous pixels with S/N$\geq$2.3 are required (corresponding to S/N$\geq$4 integrated over the full emission line). The by-eye line identification is originally performed by two different reviewers at least, with a total of 10 reviewers involved. During this phase, the same reviewers can also exclude further possible spurious detections.

Then, the spectral lines are automatically fitted with a Gaussian profile $+$ continuum. In this phase, additional emission lines, characterized by a flux below the detection threshold, can also be fit.
In the final catalog, a unique entry for each galaxy is included, and a measure of the main observational quantities is reported for each line (total flux, equivalent width (EW), FWHM). Inconsistent line identifications among different reviewers are considered in the quality flag, and only the solution associated with the line fit that minimizes the $\chi^{2}$ parameter is included in the catalog.

\subsection{Training and testing sets}
\label{SEC:cal_test_samples}
For our tests, we considered the same training and testing sets used by BA20.

In the common practice, in order to avoid overfitting problems, the training and the testing sets are independent of each other.
 In BA20, the two sets almost completely overlap (``gold'' sample). In that paper, we demonstrated both theoretically and empirically, that overfitting effects were negligible due to the design of the algorithm.

  Similarly to BA20, in order to train and test the algorithm described in this paper, we can feed \umlaut\ using a unique reference data sample without introducing overfitting effects.
  In fact, \umlaut\ tests, one by one, each of the data points of the reference sample, estimating the accuracy by averaging the results of all the tests performed. 
  Before each test, the data point under analysis is excluded from the training set and \umlaut\ is retrained from scratch, without retaining information of the previous trainings and tests.


The strategy outlined above is known as ``leave-one-out cross validation''\citep[e.g.,][]{10.2307/1267500,doi:10.1111/j.2517-6161.1974.tb00994.x,hastie01statisticallearning}. For the skeptical reader, below we explain why this choice does not generate overfitting. Additionally, the same arguments are empirically demonstrated using \umlaut, in Appendix~\ref{SECT:AppendixA}.

 Even if every data point of the reference sample is tested, this does not imply that the same data points used for training the algorithm are \emph{contemporarily} used for testing it. In fact, this is not the case, as the overlap between training and test sets is only apparent. 

  For a comparison, in common practice, a generic ML algorithm can be trained and tested by dividing a reference sample into two separate and independent subsamples. In this case, in an initial phase the first subset is used for training, while the second one is used for testing the algorithm. Then, in order to reduce the uncertainty associated with the accuracy of the algorithm, the two subsamples can be inverted to obtain an additional accuracy test. Also in this example, at the end of the process, each data point has been used for both training and testing the algorithm, but it is important to notice that none of them is used for both these two purposes at the same time. The drawback of this commonly used method is that, in each test, only a fraction of the available reference data points (about half of the sample) can be used to train the algorithm. This fact can be problematic when only a few reference data points are available.

  The ``leave-one-out'' strategy simply represents a special case of the more commonly used technique, where the training set is made of $n-1$ elements ($n$ being the total number of reference data points), while the test set is made of just one element. Additionally, instead of performing only two (or a few) tests, as when the commonly used technique is applied (by inverting training and test sets), using the ``leave-one-out'' approach, $n$ different tests are performed.

  
 In our analyses, 
  the reference sample is 
made of spectra taken in WISP fields covered by both WFC3 grisms (this choice reduces the analysis area to $\sim$900 arcmin$^{2}$).
We considered spectra with two or more lines detected (S/N$>$2$\sigma$) and consistently identified by the reviewers. In order to obtain precise measures of the flux ratios (needed to correctly identify the strongest emission in the spectra), we excluded all of the lines located at the low-sensitivity ends of the grisms' wavelength range ($\lambda<$8500\AA, $\lambda>$16700\AA, and 11000\AA$<\lambda<$11400\AA).

 For consistency with BA20, while the algorithm is tested on all the sources of the reference sample (2283 spectra), we always excluded from the training set all the sources without a valid measurement in the F110W photometric band (155 spectra).
In BA20, this information was needed to calibrate one of the modules of the supervised algorithm.
The same requirement is not necessary to train \umlaut, but in order to compare the performances of the two algorithms on the same basis, we retain this condition. 
Following the definition given in BA20, in the rest of the paper we generically refer to these reference data as the ``gold'' sample.

It is important to notice that multiple lines are detected (above 2$\sigma$) in each of the spectra used during the training and testing phases. However, \umlaut\  is always blind to the presence of additional lines except for the brightest one. Instead, for our analysis (Section \ref{SECT:Single_line_classification}), we consider spectra characterized by  similar spectral coverage (i.e., observations in both the two grisms are available) but with only one line detected.
Indeed, the aim of our analysis is to demonstrate that \umlaut\  can be successfully used to automatically classify individual emission lines detected in similar or larger spectroscopic surveys.
\newpage

\section{\umlaut: the unsupervised algorithm}
\label{SECT:THE_ALGORTHM}

\subsection{Rationale}

Our unsupervised ML algorithm for regression can be considered a variant of the $K-$nearest neighbor ($K$-NN) algorithm \citep[see, for example,][]{doi:10.1080/00031305.1992.10475879}.

 More precisely, \umlaut\ combines different methods, such as ranking, k-NN classification, and the simultaneous treatment of ranked and unranked spaces. Similar nonparametric multivariate methods can be found in the literature and span several decades \citep[see, e.g.,][and references therein]{puri1971nonparametric,hettmansperger2010robust,oja2010multivariate,wilcox2011introduction,guha2018application,Tao_Qiu_2021}.

 Being mostly based on the $K$-NN algorithm, the functioning of \umlaut\ relies on the following assumption:

\begin{enumerate}
\item\label{principle1}{\emph{Data samples that are similar in a sufficiently large amount of independent dimensions ($N>>$) tend to be similar also when considering an additional independent dimension ($N+1$).}}
\end{enumerate}

The principle just expressed is empirically demonstrated in many examples from the real world (e.g., two twins in a picture tend to be twins also in reality). On the other hand, there are also exceptions, such as optical illusions, that are often based on the impossibility of directly sensing the third dimension \citep{sugihara:LIPIcs:2018:8793} \footnote{If optical illusions were not actual \emph{exceptions}, our brain would have evolved to prevent these illusions from continually (and therefore dangerously) fooling us.
This does not mean that we are able to perceive actual reality as it is \citep[see, e.g.,][]{Hoff_article_2010}}

Exploiting principle \ref{principle1}, \umlaut\  computes the expected value assumed by a given data point along the ($N+1$)$th$ dimension, once the other $N$ dimensions are known. 
For this purpose, the value assumed along all $N+1$ dimensions must be known for the data of a reference sample with statistical properties similar to those characterizing the sample from which the analysis data point is taken.

In our experiment (see Section~\ref{SECT:WISP_TESTS}), the $N$ known parameters  correspond to the following spectrophotometric quantities: apparent magnitude and size, $J-H$ color index, equivalent width, and wavelength of the observed lines. Instead, the $(N+1)th$ dimension is represented by the spectroscopic redshift. All of these quantities are known for each of the sources in our ``gold'' sample (i.e., our reference sample). 
Given this configuration, \umlaut\  can be used to estimate the redshift of a generic source under analysis, once the other $N$ parameters are known (for the same source). In particular, the redshift is computed by combining the values of $z$ expressed by the sources in the ``gold'' sample that look more similar (under the point of view of the $N$ parameters) to the source under analysis. 
In turn, the estimated redshift (or the corresponding probability distribution function) can be used to identify the nature (\ha, \oiii, \oii) of the line(s) detected in the spectra. 

 \umlaut\  is designed to identify the position (coordinates) of a data point under analysis, with respect to a multidimensional parameter space generated by the dataset used to train the algorithm. Therefore, the most accurate results are obtained when, in the $(N+1)$-dimensional parameter space, the position of an analysis data point approximately overlap with the position occupied by some of the training data points.

  This does not prevent \umlaut\  from being used on analysis data points located outside the multi-dimensional space generated during the training. In fact, as we discuss in Section~\ref{SEC:OUTPUTS}, we included the possibility of obtaining the value of the $N+1$ parameters also outside the edges of the multidimensional parameter space (``fit'' option). However, in the current version of the algorithm, this feature is mostly meant to limit border effects, rather than to extrapolate outputs well outside the edges of the multidimensional parameter space.

 In any case, \umlaut\ fails if the analysis data point is located outside the range covered by the training set, along the $(N+1)th$ dimension (i.e., the unknown parameter). For example, if \umlaut\  is used to estimate the redshift of a source located at, e.g., $z\sim5$, but it is trained with a dataset limited to $z<2$, then the algorithm will probably return a redshift estimation close to $z\sim2$ (i.e. the upper limit reached by the data of the training set).

\subsection{Structure of the algorithm}

The rationale discussed above can be generalized as follows.
Given a data point {\bf D$^{0}$}$(d^{0}_{1},...,d^{0}_{N})$, with valid measurements $d^{0}_{i}$ along $N$ different dimensions and a reference sample [{\bf D$^{1}$}$(d^{1}_{1},...,d^{1}_{N+1})$,...,{\bf D$^{M}$}$(d^{M}_{1},...,r^{M}_{N+1})$], made of $M$ data points with valid measurements along $N+1$ dimensions (and statistically similar to the sample from which {\bf D$^{0}$} is drawn), \umlaut:
\begin{itemize}
\item{ identifies the $K$ closest reference data points {\bf D$^{j>0}$} in an $N$-dimensional space built upon the $N$-dimensional parameter space;}
\item{computes $d^{0}_{N+1}$ (the unknown parameter) by combining the measurements [$d^{1}_{N+1},...,d^{K}_{N+1}$] of the $K$ closest data points, along the $(N+1)th$ dimension.}
\end{itemize}

In general, the $N$ parameters $d_{i}$ describing a given data point can be inhomogeneous. In our test, each spectral line is associated with measurements of flux, wavelength, apparent size of the emitting source, etc. Consequently, depending on the physical units used to express the $N$ measurements, or the scale adopted to represent them (linear, logarithmic, etc.), the reciprocal distance of the $M$ data points can be completely different, with some of the $N$ parameters dominating over some others. Additionally, data can be distributed very differently along the $N$ coordinates. For example, they can be continuously distributed along some axes and show discontinuities along other axes, they can show homogeneous densities moving along each axes or not, etc. 
To overcome these problems, 
the $N$-dimensional space considered by \umlaut\  is built upon \emph{ordinal} distances. Basically, instead of considering the actual $M$ reference measurements $[d^{1}_{i},...,d^{M}_{i}]$ along a given dimension $i$, we consider their position in the ordered sample: $[p^{1}_{i},...,p^{M}_{i}]$. In other words, we convert the input measurements into rank variables.
The problematics related to the ranking are described in Section~\ref{SECT:ORDINALIZATION}


Even considering homogeneous coordinates, not all the $N$ dimensions are equally relevant when computing the expected value of {\bf D$^{0}$} along the $(N+1)th$ dimension ($d^{0}_{N+1}$). 
In fact, the statistical association between each of the $N$ parameters and the $(N+1)th$ dimension may be expressed by higher or lower correlation coefficients.
At the same time, some of the $N$ dimensions might not be necessarily independent of each other, so that the reciprocal distance of the reference data points could be dominated by the combination of the reciprocally dependent dimensions.

It is important to notice that the degree to which each of the $N+1$ parameters is in relation with the others might only have local validity in the associated ($N+1$)-dimensional parameter space. For example, the apparent magnitude is a good redshift tracer for the largest sources (i.e., a bright extended source can only be local), while it does not provide significant information when the sources are compact and faint (i.e., a faint source could either be a small local galaxy or an intrinsically bright and distant one).

In order to mitigate all of these effects, \umlaut\  locally weights each of the $N$ dimensions depending on their ability, when combined with the other dimensions, to allow for the identification of more precise values of the output parameter $d^{0}_{N+1}$. The details of the coordinate weighting are described in Section~\ref{SECT:coord_weigh}.

 Concerning the weighting of the training data points, \umlaut\ does not counterbalance the relative proportions of data belonging to different classes (or occupying different areas of the multidimensional parameter space). We discuss the details of this choice in Section~\ref{SECT:data_weighting}.

\subsubsection{Ranking coordinates}
\label{SECT:ORDINALIZATION}
Given an $N$-dimensional data point {\bf D$^{0}$}$(d^{0}_{1},...,d^{0}_{N})$, the closest $K$ reference data in the corresponding N-dimensional space can be identified by computing the Euclidean $N$-dimensional distance. Specifically, the distance $r^{j}$ of the $j$-th data point from D$^{0}$ is given by:
\begin{equation}
\label{EQ:TP1}
r^{j}=\sqrt{\sum^{N}_{i=1}{(d^{j}_{i}-d^{0}_{i})^{2}}}.
\end{equation}

Following Equation~\ref{EQ:TP1}, the distance between two data points is strongly influenced by the physical units and by the scale (linear, logarithmic, or other) used to represent the measurements along the $N$ heterogeneous dimensions.

For instance, the spectral lines in our WISP sample are associated with the wavelength at which the lines are detected ($N_{1}=\lambda_{\mathrm{obs}}$, with $8000$ \AA $<\lambda_{\mathrm{obs}}<17000$ \AA) and to the apparent infrared magnitude ($J$ and $H$ bands) of the emitting source ($N_{2}=\mathrm{mag}$, with $18<\mathrm{mag}<27$). In this simplified example ($N=2$), unless the wavelength of two different lines is very similar ($\Delta\lambda<100$ \AA), the relative $N$-dimensional distance of two points is strongly dominated by the observed wavelength ($N_{1}$). Consequently, Equation~\ref{EQ:TP1} would be simplified to $r^{j}\sim | d^{j}_{1}-d^{0}_{1} |$. Conversely, the magnitude of a source would not play any role when trying to identify the ``most similar'' data points in the $N$-dimensional space, as $| d^{j}_{1}-d^{0}_{1} |>>| d^{j}_{2}-d^{0}_{2} |$. 
On the contrary, we want to exploit \emph{all} the information provided by the $N$ different dimensions and, even more important, we do not want this information to depend on the scale or on the physical units considered.

\begin{figure*}[!ht]
\centering
\includegraphics[width=15cm]{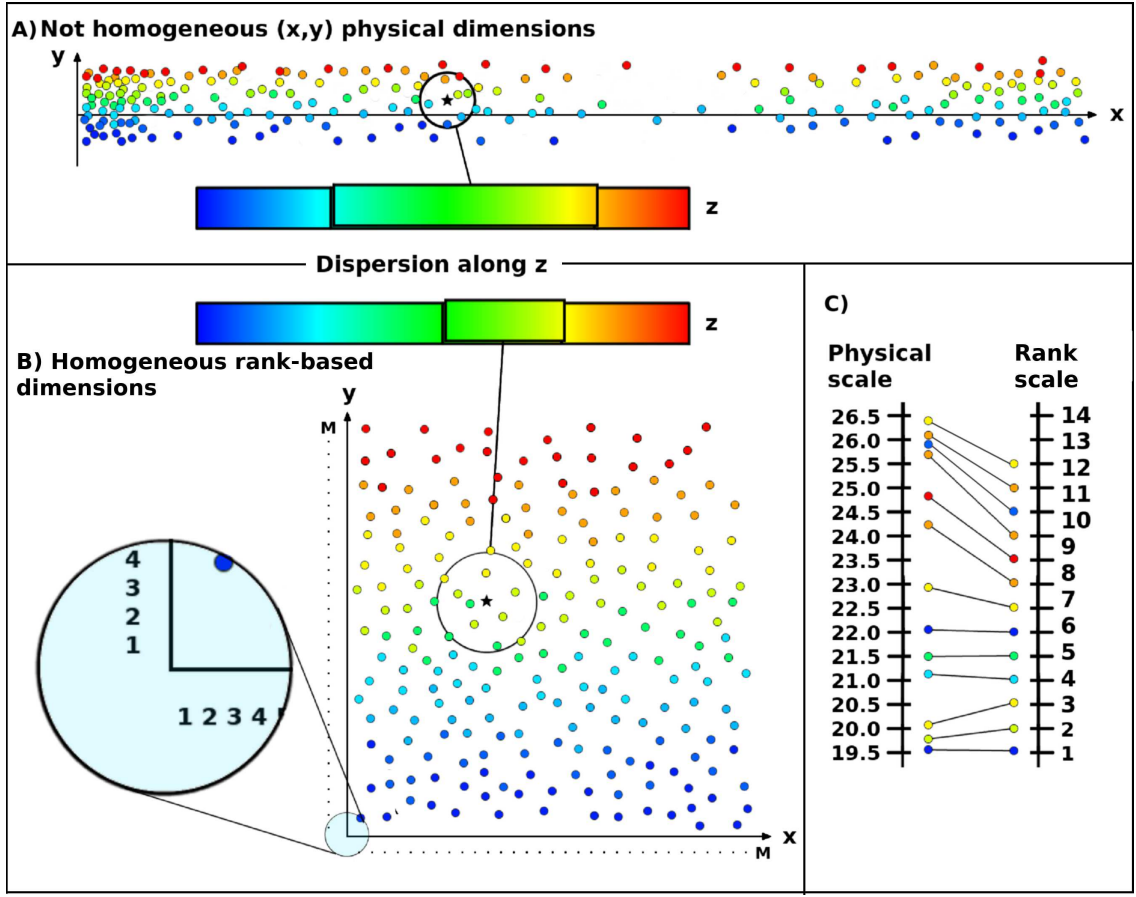}
\caption{Schematic illustration of the transformation of a generic pair of physical coordinates $x_{\mathrm{phy}}, y_{\mathrm{phy}}$ (A) into rank-based coordinates $x_{\mathrm{rank}}, y_{\mathrm{rank}}$ (B). To this purpose, every physical measurement (data points in the reference sample) along a certain coordinate is converted into a positional value along the corresponding rank scale (C). In this example, the $y$-axis is clearly correlated with $z$ (represented with a color scale), while there is no correlation between the $x$-axis and $y$ or $z$. In the original physical space, the $x$-axis clearly dominates when determining the distance between a given data point (black star symbol) and the closest data points of the reference sample (data points inside the circle). Consequently, the average $z$ value of the closest data is associated with a relevant dispersion. Using rank-based coordinates (B), the data points are positioned into a square space ($M\times M$) with homogeneous density. In the space of the rank-based coordinates, $x$ and $y$ are equally weighted when determining the distance between two data points. In this case, the average value of $z$ for close data points is associated with reduced dispersion, as the effect of the uncorrelated $x$-axis is mitigated.  }
\label{img:ORDINALIZATION}
\end{figure*}

In order to overcome the problems exposed above and equally weight the natively heterogeneous $N$ dimensions (before any successive weight strategy), \umlaut\  converts the original \emph{cardinal} coordinates into \emph{rank-based} coordinates, as schematically represented in Figure~\ref{img:ORDINALIZATION} for a simplified two-dimensional parameter space. While cardinal coordinates represent the absolute position occupied by a numerical object inside a (possibly infinite) set, rank-based coordinates denote the position occupied in the ordered sequence of elements of a subsample of the original set (possibly coincident with the entire original set).

Hence, for each dimension $i$, the data points in the reference sample are ordered from the smallest to the biggest, and a positional value $p^{j}_{i}$ is assigned to each data point $j$. 
Before any weighting, the distance between a given data point and a data point in the reference sample is computed as:
\begin{equation}
\label{EQ:TP2}
r^{j}=\sqrt{\sum^{N}_{i=1}{(p^{j}_{i}-p^{0}_{i})^{2}}}=\sqrt{\sum^{N}_{i=1}{(\Delta^{j}_{i})^{2}}}.
\end{equation}

\subsubsection{Coordinate weighting}
\label{SECT:coord_weigh}
As discussed in Section~\ref{SECT:ORDINALIZATION},  rank-based coordinates allow us to remove the dependence of the output value of $d^{0}_{N+1}$ on the arbitrary choice of physical units and scales used to express the $N$ input parameters. In fact, when identifying the closest (i.e. most similar) data of a reference sample in the $N$-dimensional parameter space, rank-based coordinates equally weight all the input parameters.

However, not all $N$ input parameters are necessarily related (or similarly related) to the $(N+1)th$ coordinate.
Uncorrelated input dimensions decrease the precision of $d^{0}_{N+1}$, as the distance between the given data point and the data points in the reference sample (Equation~\ref{EQ:TP2}) is randomly modified.
Some parameters may be well correlated with $d^{0}_{N+1}$ in some areas of the $N$-dimensional space but may be poorly correlated in others. 
In addition, some of the $N$ dimensions may be redundant or simply correlated with each other. In this case, the use of  rank variables would not prevent the overweight of the redundant input dimensions. Finally, even if a certain parameter is physically related to the output dimension, the uncertainty on the measurements may completely hide such a correlation.
  
 In order to identify the most effective set of weights, \umlaut\ follows an iterative reinforcement learning (RL) strategy \citep[see, e.g.,][]{Sutton1998}. At each iteration, the input dimensions (rank-based coordinates) are rescaled of a certain quantity. If such a scaling improves the output precision, then the input dimensions are correspondingly rewarded (or penalized). All the rewards and penalties are summed up iteration after iteration until the resulting combination of weights converges toward a local optimum.

Specifically, in each iteration, the rank-based parameter space is scaled along different directions (dimensions) and a new set of $K$ closest neighbors is found. Hence, the corresponding value of $\sigma^{0}_{N+1}$ is measured. In the successive iterations, the algorithm keeps deforming the space following the most promising gradient of $\sigma^{0}_{N+1}$. In other words, the $N$-dimensional parameter space is associated with an $N$-dimensional scaling space that is probed by \umlaut\  using the iterative procedure described.
At the end of this process, a set of weights $w_{i}$ is included in Equation~\ref{EQ:TP2} as follows:
\begin{equation}
\label{EQ:TP3}
r^{j}=\sqrt{\sum^{N}_{i=1}{[(p^{j}_{i}-p^{0}_{i})(1+w_{i})]^{2}}}=\sqrt{\sum^{N}_{i=1}{[(\Delta^{j}_{i})(1+w_{i})]^{2}}}.
\end{equation}

The iterative procedure is organized in steps, as follows:
\begin{enumerate}

\item{\label{item:start} Once the $K$ closest elements of a given data point {\bf D$^{0}$} are identified in the $N$-dimensional rank-based space, \umlaut\  computes the output parameter $d^{0}_{N+1}$ and its associated dispersion $\sigma^{0}_{N+1}$ by combining the values $d^{1}_{N+1},..., d^{K}_{N+1}$ assumed by the $K$ closest elements along the $N+1$ dimension.}

\item{\label{item:coord_def} One by one, the distances of the reference data points from {\bf D$^{0}$}, projected along each of the $N$ rank-based coordinates ($\Delta^{j}_{i}=p^{j}_{i}-p^{0}_{i}$, in Equation \ref{EQ:TP2}), are individually scaled to $\Delta^{j}_{i}(\mathrm{test})=\Delta^{j}_{i}\times(1+f*_{\mathrm{exp}})$, with $f*_{\mathrm{exp}}$ set to 1.0 for the first iteration. While scaling the $i$-th projection, all the other projections remain fixed to their original values. As a consequence, each single scaling corresponds to a new set of total distances between the data point {\bf D$^{0}$} and the reference data ($r^{j}$ in Equation \ref{EQ:TP2}).
  For each scaling, a new set of  $K$ closest elements is identified and a new pair  $(d^{0}_{N+1},\sigma^{0}_{N+1})|_{i}$ is obtained.}
  
\item{\label{item:Pen_rew} The dispersion associated with the $i$-th scaling ($\sigma^{0}_{N+1}|_{i}$) is compared with the dispersion obtained from the K closest data points identified before the scaling (i.e., before point~\ref{item:coord_def}).
After scaling all the $N$ dimensions, each projected distance $\Delta^{j}_{i}$ is replaced by $\Delta^{j}_{i}\times(1+w_{i})$, with $w_{i}$ initially set to zero and increased by: 
  \subitem{$\Delta w_{i}=+0.2f*_{\mathrm{exp}}$, if $\sigma^{0}_{N+1}|_{i}<\sigma^{0}_{N+1}$;}
  \subitem{$\Delta w_{i}=-0.1f*_{\mathrm{exp}}$, if $\sigma^{0}_{N+1}|_{i}\geq\sigma^{0}_{N+1}$;}
  \subitem{$\Delta w_{i}=0$, if $\sigma^{0}_{N+1}|_{i}<\sigma^{0}_{N+1}$ and $w_{i}=-1.0$.}

  In other words, the parameter $\Delta w_{i}$ rewards or penalizes the test scaling along dimension $i$ depending on the precision measured. When $\Delta w_{i}$  acts like a penalty ($\Delta w_{i}<0$), the input dimension $i$ is contracted as a result of the test. After some iterations, the same dimension can be ruled out if $w_{i}$ reaches the value of -1.
As $w_{i}$ is initially set to $w_{i}=-0.1f*_{\mathrm{exp}}$, the condition $w_{i}=-1$, corresponding to $\Delta^{j}_{i}=0$ (i.e., input dimension uncorrelated with the output parameter), can be obtained only after completing 10 iterations at least.}

\item{\label{item:spec_rew_and_f} The coordinate along which the scaling provided the lowest value of $\sigma^{0}_{N+1}|_{i}$ is further scaled by a factor $\Delta w_{i}=+0.4f*_{\mathrm{exp}}$, unless none of the $N$ scaling generated a value of $\sigma^{0}_{N+1}|_{i}$ smaller than the reference $\sigma^{0}_{N+1}$. In the latter case, the scaling factor $f*_{\mathrm{exp}}$ is reduced to $f*_{\mathrm{exp}}/2.0$   }

\item{The process is reiterated from point~\ref{item:start}, but now considering the new projected distances $\Delta^{j}_{i}\times(1+w_{i})$ in place of the original ones and the new scaling factor $f*_{\mathrm{exp}}$, if required. When a maximum number of iterations (set by the user) is reached, or when the dispersions $\sigma^{0}_{N+1}|_{i}$ do not improve further (i.e., $f*_{\mathrm{exp}}<0.01$), the iterations end, and the projected distances resulting from the process are considered for the final identification of the $K$ closest data points.}
  
\end{enumerate}


At the end of the process, the data point under analysis ({\bf D$^{0}$}) is associated with a position in the rank-based parameter space ($N$ coordinates) and with another position in the scaling space ($N$ additional coordinates). The position in the scaling space represents the matrix of multiplicative factors required by each of the $N$ coordinates of the parameter space in order to minimize the dispersion on the output parameter $\sigma^{0}_{N+1}$.

 Up to a certain point, the amplitude of rewards and penalties (defined at step \ref{item:Pen_rew}) influences the number of iterations required for the algorithm to converge (The smaller the steps, the higher the number of iterations required). In any case, these quantities automatically scale down during the iterative process. This is due to the halving of $f*_{\mathrm{exp}}$ (see step \ref{item:spec_rew_and_f}). As a consequence, once some generic ``rules of thumb'' are followed, the initial values of penalties and rewards do not sensibly modify the output accuracy.
  
 As a first ``rule of thumb'', the number of iterations and the initial penalty should be set so that $N_{\mathrm{iter}} > |1/\mathrm{penalty}|$. Here, the penalty corresponds to the initial value assumed by $\Delta w_{i}$ when the precision gets worse from an iteration to another (second among the three cases listed in step \ref{item:Pen_rew}). This rule should provide the algorithm with a sufficient number of iterations to rule out one of the input parameters when it does not correlate at all with the output (i.e., it allows $w_{i}$ to reach $-1$ before all the maximum number of iterations is reached).

On the other hand, a too-large initial penalty, such as $\Delta w_{i}=-1$, could rule out one of the input dimensions after the first iteration, i.e., when the multidimensional surface has not been properly sampled yet. Similarly, too-large rewards would move the current position on the multidimensional surface beyond the radius explored (the value of $\Delta^{j}_{i}(\mathrm{test})$ defined in step \ref{item:coord_def}). It follows the second ``rule of thumb'': the initial scaling should not exceed 50\%-100\% of the initial value (i.e., $\Delta w_{i}<0.5-1$).

Following the rules indicated above, the most effective strategy to set the initial values of rewards and penalties is to run a series of tests, starting with large values of $|\Delta w_{i}|$ ($|\Delta w_{i}|\sim 0.5-1$) and allowing for a few iterations (small $N_{\mathrm{iter}}$). At this point, the value of  $N_{\mathrm{iter}}$ should be increased until the accuracy stabilizes around a sort of ``plateau''. Then, penalties and rewards can be decreased by a certain quantity ($\sim0.1$), and $N_{\mathrm{iter}}$ increased until a new ``accuracy plateau'' is found. This process can be reiterated until each new ``accuracy plateau'' guarantees more accurate results than the previous one. On the contrary, the process should be stopped, as there is no sense in increasing the number of iterations without an improvement in the output precision.

The strategy outlined does not guarantee that the absolute minimum of $\sigma^{0}_{N+1}$ is reached at the end of the process, but it allows \umlaut\  to get close to a local minimum.
In order to determine the position of the absolute minimum, the algorithm should probe every location of the entire scaling space (i.e., every possible scaling factor along each coordinate and every combination of scaled coordinates). However, this is computationally infeasible, especially when the number of dimensions considered becomes large.

\begin{figure*}[!ht]
\centering
\vspace*{2mm}
\includegraphics[width=16cm]{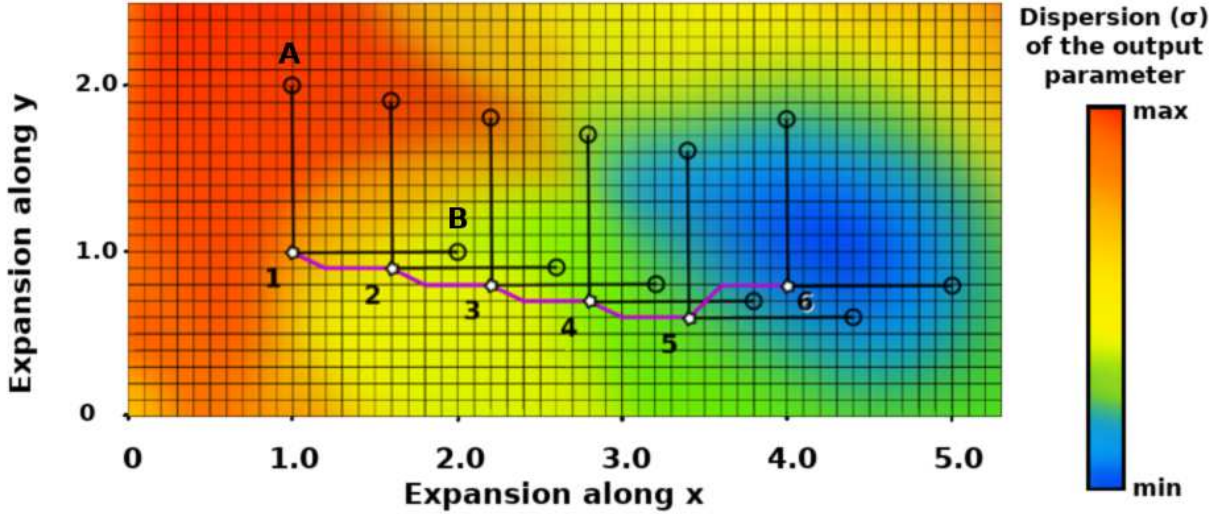}
\vspace*{2mm}
\caption{ Simplified two-dimensional projection of the iterative process used by \umlaut\  to weight two out of $N$ input parameters $(x,y)$ (in this example, the additional dimensions are not scaled. The plot represents a portion of the scaling space $(x',y')$ probed by \umlaut\  during six different iterations (star symbols numbered from 1 to 6). Each $(x',y')$ position identifies a differently scaled parameter space, with the (1, 1) position corresponding to no scaling. The color scale represents the dispersion $\sigma^{0}_{N+1}$ associated with the output parameter $d^{0}_{N+1}$, as it would result by scaling the parameter space as indicated by the $(x',y')$ coordinates in the scaling space. The value of $\sigma^{0}_{N+1}$ in each position of the scaling space can not be analytically computed but it can be obtained only numerically by scaling the parameter space correspondingly. For this reason, $\sigma^{0}_{N+1} (x',y')$ is not known \emph{a priori}. The iteration process starts from position (1, 1) in the scaling space. At the beginning of each iteration (star symbols in the plot), \umlaut\  identifies the $K$ closest data points of the reference sample, in the corresponding parameter space. These data are used to estimate the output parameter $d^{0}_{N+1}$ and its associated dispersion $\sigma^{0}_{N+1}$.
  Hence, the projections of the distances of the reference data points along each of the $N$ dimensions of the parameter space are scaled by a factor $1+f*_{\mathrm{exp}}$ (horizontal and vertical segments starting from the reference positions of each iteration), with $f*_{\mathrm{exp}}$ initially set to 1.0. A new set of $K$ closest data points is identified in each of the corresponding modified parameter spaces (for the first iteration, A and B positions in the expansion space). Then, $N$ pairs $(d^{0}_{N+1},\sigma^{0}_{N+1})$ are estimated. The dispersion $\sigma^{0}_{N+1}$, computed in the $N+1$ positions of the scaling space (star, A and B points in this example) is compared, and the parameter space is finally scaled approximately following the decreasing gradient of $\sigma^{0}_{N+1}$ (the purple line connecting the star symbols). At this point, a new iteration starts. The trajectory (purple lines) is defined by the rewards and penalties defined in step 3 and 4 of the iterative process (Section \ref{SECT:coord_weigh}). In the example shown in this figure, we considered $\Delta w_{i}$=-0.1 (penalty), $\Delta w_{i}$=0.2 (reward), and $\Delta w_{i}$=0.4 (special reward). }
\label{img:exp_stps}
\end{figure*}

In the current configuration, $N+1$ positions in the scaling space are tested by \umlaut\  during each iteration. For each of these positions, the algorithm determines the K closest elements in the reference sample and computes the pair $(d^{0}_{N+1},\sigma^{0}_{N+1})$. The situation described is graphically represented in Figure~\ref{img:exp_stps}.

Thanks to the iterative strategy discussed in this section, \umlaut\  can provide more precise and accurate outputs, especially in the case of strong variations of the degree of correlation between the input parameters and the output dimension.
 Noise introduced by the input parameters that do not correlate \emph{at all} with the output can not be completely suppressed. As shown in Appendix B, accuracy degrades significantly when a relevant fraction (1/2 to 2/3) of the input parameters are unrelated to the output. However, this degradation is limited when the weighting strategy is activated.
This fact is of particular importance when there is no \emph{a priori} knowledge of the degree of correlation between input and output.

 As a final remark, it should be noted that the weighting strategy described in this section implicitly takes into account the uncertainties associated with the input parameters. In fact, greater uncertainties imply lower degrees of correlation with the output and, therefore, lower associated weights.

\subsubsection{Weighting of the data in the training set}
\label{SECT:data_weighting}
       Besides the local weighting of the coordinates of the parameter space, outlined in Section~\ref{SECT:coord_weigh}, \umlaut\ does \emph{not} provide the possibility to weight or counterbalance the different proportions of training data belonging to different classes.
      
       On the one hand, this feature may decrease the accuracy in the regions of the $N$-dimensional parameter space characterized by degeneracy (with respect to the output parameter). In these regions, overrepresented subsamples of the training set will implicitly receive higher weights. However, in the case just described, the algorithm would not return accurate results anyway, due to the intrinsic degeneracy among the input parameters. On the other hand, where the degeneracy among the input parameters is limited, then different populations occupy different locations of the $N$-dimensional parameter space. In this case, the effects of a weighting strategy on the outputs would be limited.

      This argument can be graphically understood by observing Figure~\ref{img:ORDINALIZATION} (bottom panel). In this example, the input parameter $x$ does not provide any information on the output $z$, but the degeneracy between $y$ and $z$ is limited. In other words, the parameter $y$ allows us to break the degeneracy with respect to $z$. As a consequence, data points belonging to different classes (high-z sources in red, versus low-z sources in blue) occupy different locations of the bidimensional parameter space, and this allows \umlaut\ to determine $z$ with acceptable accuracy.  In this context, if we were artificially increasing the proportion of high-$z$ sources with respect to the low-$z$ ones, this modification would not dramatically change the resulting estimation of $z$, as the additional data points would locate themselves accordingly to the classes to which they belong. Under these modified conditions, the properties of the nearest data points would not differ substantially from those characterizing the old set of neighbors.

 The absence of a weighting feature allows \umlaut\ to provide more significant output probability indicators when training and analysis sets are sampled from the same survey (or from surveys with similar characteristics). In fact, when the inputs can not provide significant information on the output parameter (i.e. there is degeneracy), then the probability indicators must reflect the relative proportion of the different classes considered (i.e., a randomly selected object is more probably a common object than a rarer one).

 The feature just described does not prevent the user from creating samples of rare objects even when some degeneracy among the input parameters is present. For example, the output PDF (see Section~\ref{SEC:OUTPUTS}) can be used to select data points characterized by values of probability, in particular ranges of the output dimension, exceeding a certain threshold set by the user. Obviously, the choice of such a threshold would be a compromise between contamination and completeness, depending on the actual level of degeneracy.

\subsubsection{Selection of the closest data points}
\label{SECT:K_CLOSE}
After the conversion of the cardinal coordinates into rank variables and the successive weighting phase, \umlaut\  identifies the final set of $K$ closest (i.e., more similar) data in the training sample.
The $K$ parameter can be set to a fixed value by the user, or alternatively, it can vary inside a preselected range. In the latter case, the algorithm automatically selects the value of $K$ that minimizes the dispersion associated with the output parameter ($\sigma^{0}_{N+1}$).

While there is not a precise rule to follow when setting the value of $K$, some general principles should be taken into account. In the first instance, a statistically significant amount of data ($K\gtrsim 10$) should be considered. On the other hand, considering too many reference data could lead to a less precise estimate of $d^{0}_{N+1}$ (i.e. to higher values of $\sigma^{0}_{N+1}$), as data that are particularly distant in the rank-based parameter space (i.e., data that are not similar to each other) are taken into account in the computation. In this sense, $K$ should not represent a relevant fraction of the reference sample ($K\lesssim 5\%$ of the entire reference sample).

\subsubsection{Outputs}
\label{SEC:OUTPUTS}

The most important output of \umlaut\  is the value of the unknown parameter $d^{0}_{N+1}$, estimated for the input data point {\bf D$^{0}$}.
After selecting the $K$ closest elements in the reference sample, $d^{0}_{N+1}$ can be obtained in different ways:
\begin{itemize}
\item{as the average of the $K$ values $d^{j}_{N+1}$ (for $j=1,...,K$);}
\item{as the median of the $K$ values $d^{j}_{N+1}$ (for $j=1,...,K$);}
\item{as the linear fit of the $K$ data {\bf PD$^{j}=$}$\{p^{j}_{1},...,p^{j}_{N},d^{j}_{N+1}\}$ in the rank-based $(N+1)$-dimensional space (multiple linear regression). Among the various possibilities presented in this list, this option is the one that minimizes border effects due to analysis data points located at the edge (or outside) the borders of the multidimensional parameter space defined by the training dataset};
\item{as the weighted average of all the values $d^{j}_{N+1}$ in the reference sample. The weight associated with each value $d^{j}_{N+1}$ corresponds to the value assumed by the Gaussian function defined as $y=e^{-\frac{1}{2}(x/\sigma)^{2}}$ and with $\sigma=K$, in $x=j$. In this context, after ordering the distances of the data in the reference sample (Equation~\ref{EQ:TP3}), $j$ represents the position of the $j$-th closest data point with respect to {\bf D$^{0}$}. Differently from the previous possibilities, when computing the output using this option, the entire training set is taken into account (the $K$ closest data points just have the most relevant weight)}; 
\item{from the probability distribution function (PDF) provided by \umlaut. The PDF is obtained from the smoothed histogram of the values $d^{j}_{N+1}$, with $j=0,...,M$, weighted as discussed in the previous point. In this case the user can autonomously decide how to obtain $d^{0}_{N+1}$ from the PDF itself.  Similarly to the weighted average option, also in this case all the training data points are taken into consideration.}
  
\end{itemize}
Besides estimating the unknown parameter $d^{0}_{N+1}$, \umlaut\  provides the associated uncertainty $\sigma^{0}_{N+1}$. To this purpose, $\sigma^{0}_{N+1}$ can be computed in different ways as well:
\begin{itemize}
\item{as the actual standard deviation of the $K$ values $d^{j}_{N+1}$ (for $j=1,...,K$);}
\item{considering the 16th ($d^{16}_{N+1}$) and 84th ($d^{84}_{N+1}$) percentiles of the distribution of the $K$ values $d^{j}_{N+1}$ (for $j=1,...,K$). In this case, $\sigma^{0}_{N+1}$ is obtained as the maximum between $\langle d_{N+1}\rangle - d^{16}_{N+1}$ and  $d^{84}_{N+1} - \langle d_{N+1}\rangle$, with $\langle d_{N+1}\rangle$ corresponding to the median $d^{j}_{N+1}$;}
\item{as the average between the two previous solutions;}
\item{from the probability distribution function (PDF) provided by \umlaut. In this case the user can autonomously decide how to obtain $\sigma^{0}_{N+1}$ from the PDF itself. }
\end{itemize}

As described above, both $d^{0}_{N+1}$ and $\sigma^{0}_{N+1}$ can be computed from the output PDF. In order to estimate this PDF, \umlaut\  considers all the data in the training sample, each of which is associated with a value of the output parameter $d^{j}_{N+1}$. These values are put in a histogram where each element is weighted based on the corresponding distance of reference and analysis data points, computed in the rank-based parameter space. To this purpose, the weight associated with each value $d^{j}_{N+1}$ is the value assumed by the Gaussian function $y=e^{-\frac{1}{2}(x/\sigma)^{2}}$ (with $\sigma=K$), for $x=j$.

\section{The WISP survey as a test bench}
\label{SECT:WISP_TESTS}

The unsupervised approach of \umlaut\  ensures the possibility of using our algorithm in a very wide range of different contexts. Just as an example, in this paper we test \umlaut\  on the identification of single spectral lines detected in grism spectra. The results of another analysis, performed by applying \umlaut\  to linguistic data \citep{10.3389/fpsyg.2019.01528}, will be provided in a future paper (I. Baronchelli \& F. Cognola 2021, in preparation).

Concerning the identification of spectral lines, as an initial test, we compared the results obtained using \umlaut\  with those reported in BA20. The details of this analysis are discussed in Section~\ref{SECT:UNSUP_ALG_ON_WISP}.
As an additional test, we combined the unsupervised approach followed by \umlaut\  with the supervised algorithm of BA20. The results of this second analysis are detailed in Section~\ref{SECT:SUP_UNSUP_COMB}.

For both tests, we trained and tested the algorithms on the same WISP ``gold'' sample used in BA20 (see Section~\ref{SEC:DATA}).
During the training phase, the steps described in Sections~\ref{SECT:ORDINALIZATION}, \ref{SECT:coord_weigh}, and \ref{SECT:K_CLOSE} are followed. 
For the testing phase, where we measure accuracies and uncertainties, the same WISP gold sample used during the training is considered. As explained in Section~\ref{SEC:cal_test_samples} and demonstrated in Appendix~\ref{SECT:AppendixA}, this solution does not introduce overfitting problems. In fact,
\umlaut\  actually retraines itself every time a new input is tested, while the same data point is excluded from the training set. 

All of the lines in the gold sample are real and additional spectral features (to which both \umlaut\  and the BA20 algorithm are blind) are always detected in each of these spectra, allowing for their secure identification. All the accuracies computed in our tests refer to real lines as the training set does not include false identifications.

\subsection{Unsupervised approach: UMLAUT}
\label{SECT:UNSUP_ALG_ON_WISP}

In order to test \umlaut\  on the WISP gold sample, we compared the classification of the strongest emission line in each spectrum, as obtained from the output z-PDF, with the original WISP classification. As in the current configuration, the algorithm is designed for regression purposes, it does not directly classify the strongest emission line. However, 
given the wavelength of the brightest line, the output $z$-PDF provides a probability indicator for all the species/transitions considered in our tests (\ha, \oiii\ and \oii). 
Hence, the line is identified as the species/transition to which the algorithm assigns the highest probability. 

\begin{figure*}[!ht]
\centering
\includegraphics[width=5.7cm]{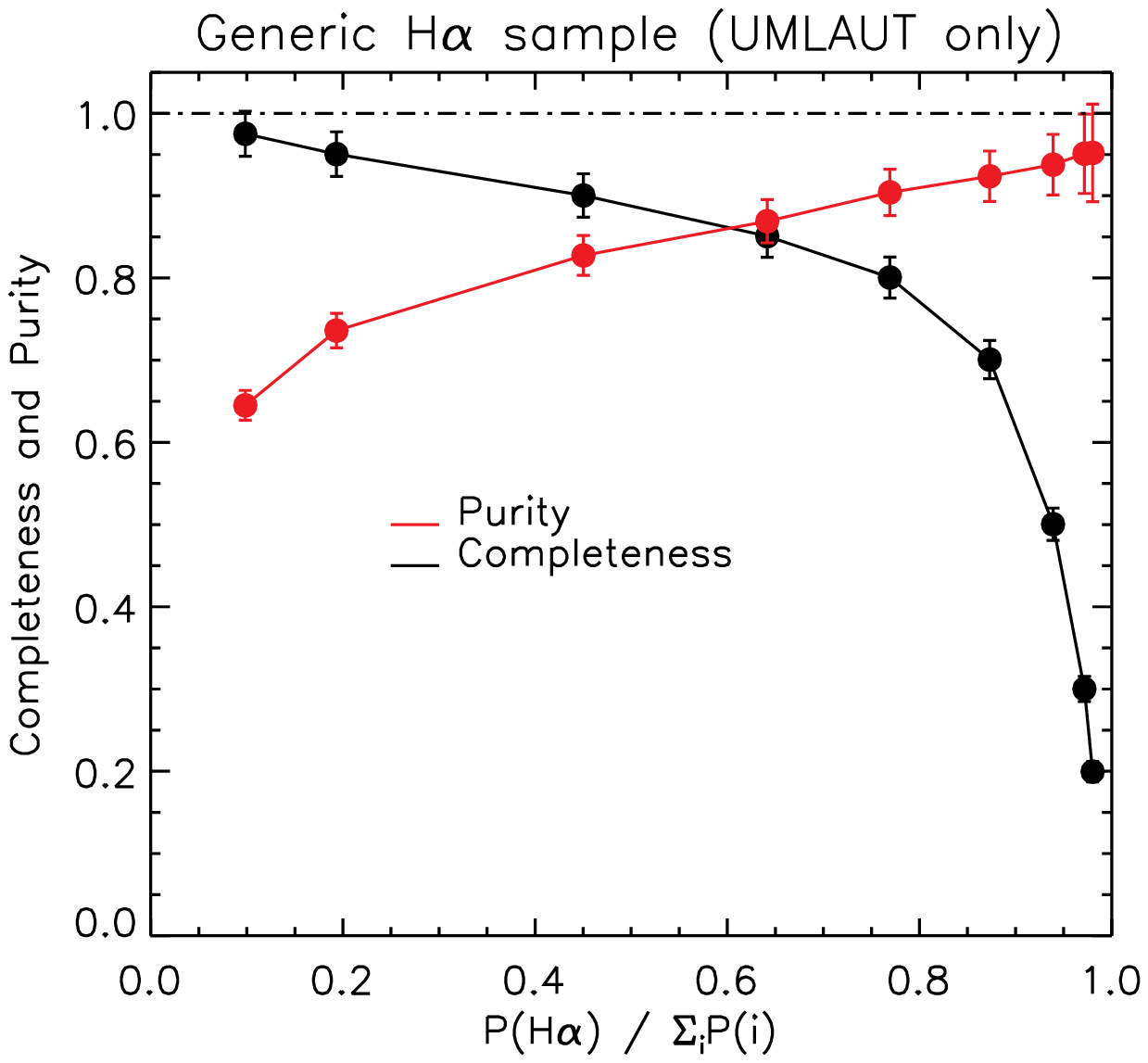}
\includegraphics[width=5.7cm]{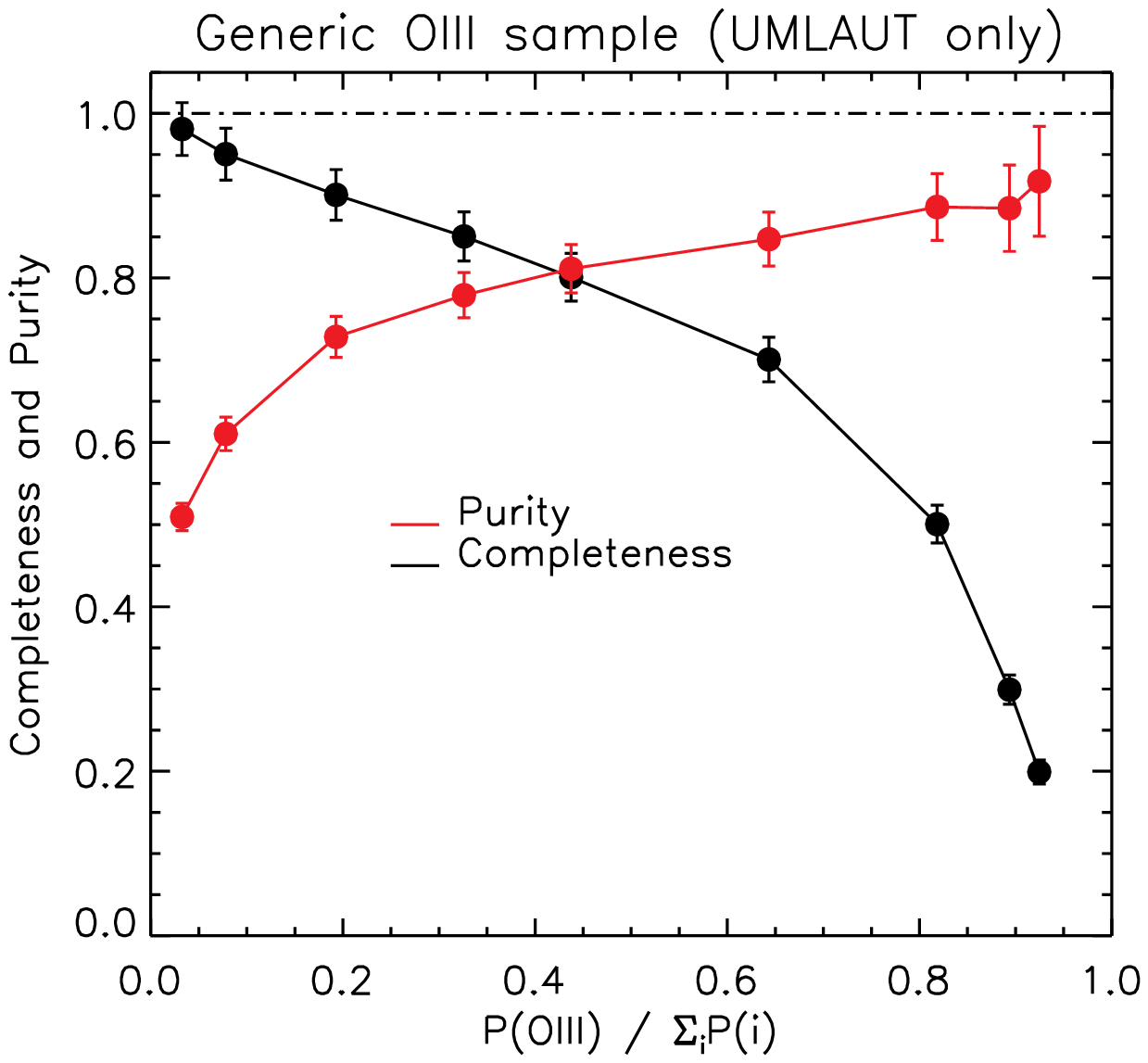}
\includegraphics[width=5.7cm]{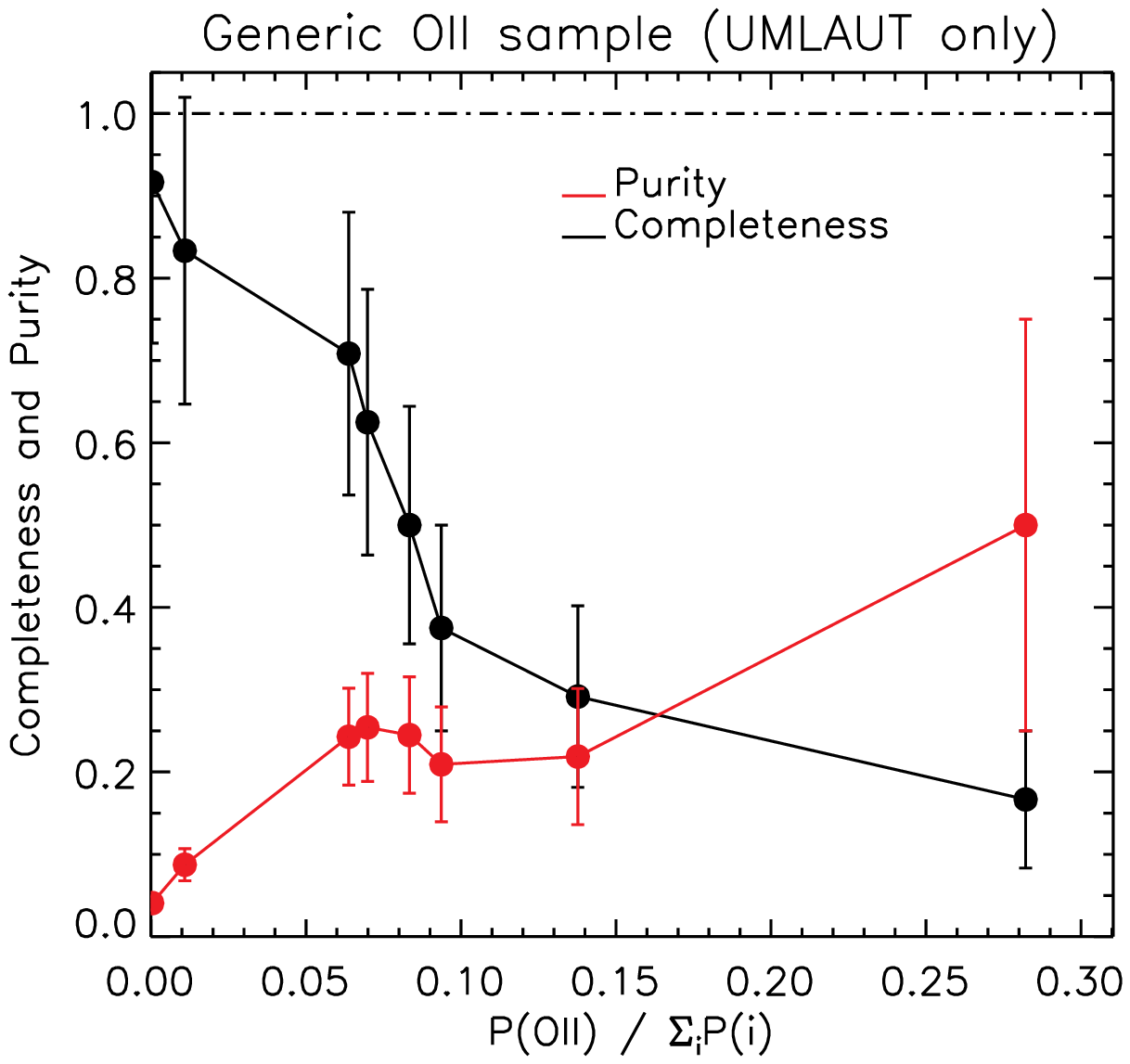}
\caption{For every data point under analysis and for each of the classes considered, we compute one probability indicator from the output PDF provided by \umlaut: P(\ha), P(\oiii), P(\oii). In our analysis, each line is associated with the label (\ha, \oiii, \oii) characterized by the highest value of the corresponding probability indicator. This approach guarantees the highest overall accuracy, but it is not the only approach that can be adopted. In these plots we show how the probability indicators (computed from the outputs of \umlaut) can be fine-tuned to select more or less complete/pure samples of sources dominated by the \ha\ ({\bf left panel}), \oiii\ ({\bf central panel}) and \oii\ ({\bf right panel}) emission. In these examples, P(\ha), P(\oiii) and P(\oii) are normalized to the sum of the probabilities computed for all the species considered ($\sum_{i}$P(\emph{i})). The error bars correspond to Poissonian uncertainties.}
\label{img:PROB_VS_PRECISION1}
\end{figure*}

 The approach just described allows us to maximize the overall accuracy (i.e., considering all the sources, regardless of their actual classification), but it is not the only possible one. Figure~\ref{img:PROB_VS_PRECISION1} shows how, by setting the selection thresholds at different levels of P(\ha), P(\oiii), or P(\oii), we can obtain more complete (or more pure) samples of the corresponding classes. This different approach is particularly useful when the overall accuracy is not the primary goal of the classification process. For example, a user may want to create a very pure sample of rare objects (regardless of the completeness). On the opposite, sometimes an initial rough automatic classification may be required in order to eliminate most of the contaminants, before implementing a more refined strategy. In this second case, the selection should maximize the level of completeness.

Besides using the same training and testing sets considered in BA20, the comparison between the performance of our unsupervised algorithm and that of BA20 is made easy by using similar input information. In particular, for each spectrum we provide \umlaut\  with the following parameters (when available):
\begin{itemize}
\item{apparent size of the emitting source, estimated from $J$ band images (the $H$ band is used when $J$ is not available);}
  \item{apparent $J$ magnitude of the emitting source (the $H$ band is used when $J$ is not available);}
  \item{$J-H$ color index;}
  \item{apparent equivalent width of the strongest emission line;}
  \item{observed wavelength of the strongest emission line;}
\end{itemize}
As shown in BA20, all these input parameters are related to the output (i.e., the redshift $z$ that we want to estimate). In this sense, in our analysis we do not need to deal with the possible presence of input parameters that do not correlate with the output. In Appendix~\ref{SECT:AppendixB}, we show how \umlaut\  would still perform well in such a case.

The main configuration parameters that we set for our tests are summarized in Table~\ref{tbl:ALG_PARAM}.
In this configuration, \umlaut\  is free to find the most effective value of $K$ inside the range $15<K<25$. We remind that, considering the output PDF, all the reference data are considered (i.e. not only the closest $K$ elements). However, the weight associated to each data point decreases when more and more distant data in the rank-based parameter space are considered (see description in Section~\ref{SECT:K_CLOSE}).

 \begin{deluxetable}{lcl}
 \tabletypesize{\footnotesize}
 \tablecolumns{2}
 \tablewidth{0pc}
 \tablecaption{Configuration of \umlaut\  during the tests}
\tablehead{Parameter & Configuration}
\startdata
 \label{tbl:ALG_PARAM}
Type of output & PDF \\
$z$-PDF step & 0.1 \\
Min. possible value of $K$ & 15 \\
Max. possible value of $K$ & 25 \\
 \enddata
 \end{deluxetable}

\subsubsection{Accuracy and uncertainties}

Similarly to BA20, we measure the accuracy (fraction of correctly classified lines) by comparing the classification of the strongest emission lines performed by \umlaut\  on gold sample spectra with the original WISP classification. Our unsupervised approach correctly classifies the strongest emission lines with an overall accuracy (83.2\%) that is comparable to that reached in BA20 (82.6\%).

In Table~\ref{tbl:ACCU_TEST_1} we report completeness and accuracy measured on three subsamples of the WISP gold sample. In each of these sub-samples, the strongest emission line is due to a different species/transition: \ha, \oiii, or \oii.
Again, with the exception represented by the \oii\ sample, the results of our test are very similar to those reported in BA20 (see Table 3 therein). 

The \oii\ sample recovered by \umlaut\  is far less complete, although significantly purer, than that in BA20 (completeness=12.5\%, versus 50\% in BA20; purity=66.7\%, versus 26.7\% in BA20). These differences can most probably be attributed to the small dimension of the \oii\ sample that makes the classification particularly sensitive to the learning strategy of the algorithm used.

 \begin{deluxetable*}{cclllllll}
 \tabletypesize{\footnotesize}
 \tablecolumns{9}
 \tablewidth{0pc}
 \tablecaption{Accuracy obtained using only \umlaut, the unsupervised algorithm described in this paper.}

\tablehead{ species/ & $\lambda_{\mathrm{obs}}$ & N$_{\mathrm{TGS}}$\tablenotemark{[a]} & N$_{\mathrm{I}}$\tablenotemark{[b]} & N$_{\mathrm{CI}}$\tablenotemark{[c]} & N$_{\mathrm{WI}}$\tablenotemark{[d]} & Completeness  & Contamination & Accuracy (Purity) \\
        Transitions  & [\AA] & & & & & (N$_{\mathrm{CI}}$/N$_{\mathrm{TGS}}$) & (N$_{\mathrm{WI}}$/N$_{\mathrm{I}}$) & (N$_{\mathrm{CI}}$/N$_{\mathrm{I}}$) }
 \startdata
 \label{tbl:ACCU_TEST_1}
 \bf{H$\alpha$}  & \bf{6564.5}          & \bf{1293} & \bf{1379} & \bf{1158} & \bf{221} & \bf{89.6\%}  &  \bf{16.0\%} & \bf{84.0\%} \\
 \bf{\oiii} & \bf{4960.3 - 5008.2} & \bf{949}  & \bf{900}  & \bf{738}  & \bf{162} & \bf{77.8\%}  &  \bf{18.0\%} & \bf{82.0\%} \\
 \bf{\oii}\tablenotemark{[e]} & \bf{3727.1 - 3729.9}  & \bf{24}   & \bf{3}   & \bf{2}   & \bf{1} & \bf{12.5\%}  &  \bf{33.3\%} & \bf{66.7\%} \\
 \hline
 \bf{Overall accuracy:} & \multicolumn{8}{l}{\bf{83.2\%}\tablenotemark{[f]}} \\
 \enddata
\tablenotetext{a}{ Number of sources, in the test ``gold'' sample (TGS), showing this transition as the most prominent emission line.)}
\tablenotetext{b}{ Number of sources, in the test ``gold'' sample, for which the algorithm associates to the possibility of this transition to be the responsible for the most prominent emission line measured, the highest relative probability (Identified, I).}
\tablenotetext{c}{ Number of sources, in the test ``gold'' sample, for which the strongest emission due to this transition is correctly identified (CI).}
\tablenotetext{d}{ Number of sources, in the test ``gold'' sample, for which the strongest emission is mistaken for this transition (wrongly identified, WI).}
\tablenotetext{e}{ Due to the small amount of data included in the training and testing sets, the precision estimated should be considered an upper limit to the actual value.}
\tablenotetext{f}{ The ``gold'' sample includes a few lines that are not originally identified as \ha, \oiii, or \oii. Because the algorithm can not be properly trained on these sets, we do not report the corresponding accuracies in this table.}
 \end{deluxetable*}

\subsection{Combining supervised and unsupervised strategies}
\label{SECT:SUP_UNSUP_COMB}
As discussed in Section~\ref{SEC:OUTPUTS}, \umlaut\  provides a probability distribution function associated with the unknown parameter (a redshift PDF in our test). On the other hand, the supervised algorithm of BA20 allows the user to integrate additional PDFs and to combine different sources of information.

\begin{figure*}[!ht]
\centering
\includegraphics[width=16cm]{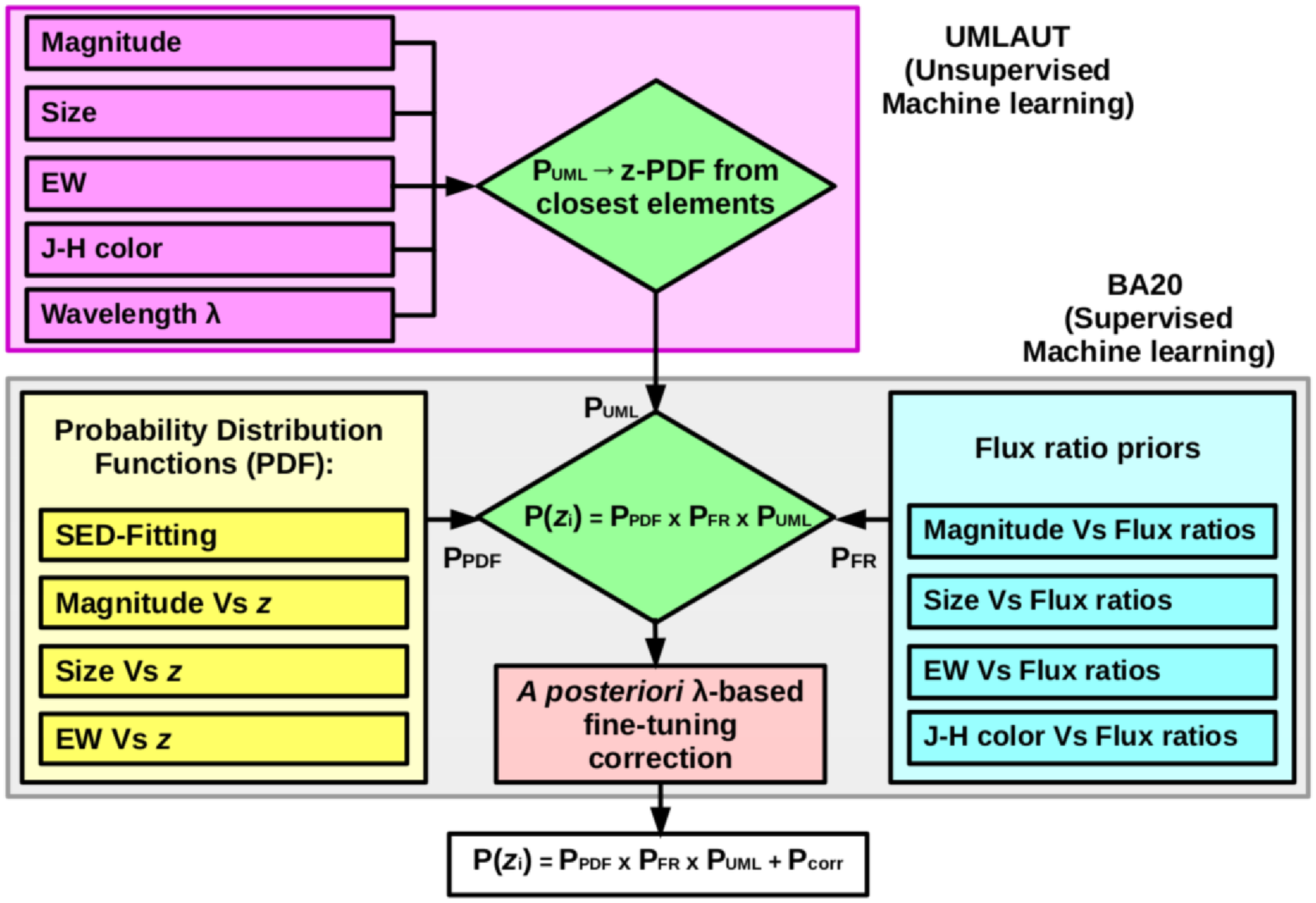}
\caption{Block diagram explaining the combination of our unsupervised ML algorithm (purple shaded area) with the BA20 supervised algorithm (gray shaded area). The BA20 algorithm is organized into three blocks. For each spectrum, the probability distribution functions block (yellow rectangle) provides a continue estimation of P($z$) between $z=0$ and $z=3.3$ (regression). This PDF is obtained using a supervised approach. The line flux ratios block (cyan rectangle) estimates which among \ha, \oiii, and \oii is more likely responsible for the strongest measured line (classification). The $\lambda$-dependent \emph{a posteriori} correction (pink rectangle) is used to fine-tune the output probability ratios. \umlaut\  outputs an independent redshift probability distribution function: P$_{\mathrm{UML}}(z)$. This $z$-PDF, obtained through an unsupervised approach, is inputted into the BA20 algorithm, before the computation of the fine-tuning correction. }
\label{img:scheme1}
\end{figure*}

We exploited these characteristics of the two algorithms to combine the two different approaches as schematically shown in Figure~\ref{img:scheme1}. It is worth noting that most of the inputs are redundant among the different blocks. 
Therefore, what differs is mostly \emph{how} they elaborate the input information.

In BA20, after combining the various sources of information, a $\lambda$-based \emph{a posteriori} prior is applied (BA20, Section 3.2.6 and Figures 11 and 12 therein).
For each line $i$ ($i$=\ha, \oiii\ or \oii), the BA20 algorithm outputs an indicator P$_{i}$ which represents the probability that the spectral feature observed is due to the line $i$ considered. At the end of the training phase, the output is compared with the expected results and a set of corrections is applied to the ratios P$_{\mathrm{H}\alpha/\mathrm{[OIII]}}$=P$_{\mathrm{H}\alpha}/$P$_{\mathrm{[OIII]}}$ and P$_{\mathrm{[OIII]}/\mathrm{[OII]}}$=P$_{\mathrm{[OIII]}}/$P$_{\mathrm{[OII]}}$ initially estimated.
This backpropagation feature allows the user to further improve the initial calibration, but the correction must be properly set each time a new source of information is included (or when a dataset with different statistical properties is considered).
Consistently with Tables 1 and 2 in BA20, in Tables~\ref{tbl:Empirical_corr_Ha_OIII} and \ref{tbl:Empirical_corr_OIII_OII} we report the corrections applied in our test to P$_{\mathrm{H}\alpha/\mathrm{[OIII]}}$ and P$_{\mathrm{[OIII]}/\mathrm{[OII]}}$, respectively.

\begin{deluxetable}{lr}
\tabletypesize{\footnotesize}

\tablecolumns{2}
\tablewidth{0pc}
\tablecaption{Combination of \umlaut\  and BA20 algorithms: Fine-tuning of the P(H$\alpha$)/P(\oiii) ratio \tablenotemark{[a]}}
\tablehead{\colhead{$<\lambda_{\mathrm{obs}} [\emph{\AA}]>$} & \colhead{$\log$P$^{\mathrm{best}}_{\mathrm{H}\alpha/\mathrm{[OIII]}}(\lambda_{\mathrm{obs}})$} }
\startdata
\label{tbl:Empirical_corr_Ha_OIII}
8763.10 & 0.439  \\
9259.92 & -0.532 \\
9771.42 & -0.329 \\
10233.7 & -0.688 \\
10730.2 & 0.199  \\
11248.4 & 1.398  \\
11755.0 & 0.849  \\
12234.3 & 0.510  \\
12755.0 & -0.167 \\
13267.0 & -0.151 \\
13767.4 & 0.403  \\
14242.6 & 0.619  \\
14721.9 & 0.403  \\
15250.4 & -0.247 \\
15743.8 & -0.151 \\
16240.9 & 0.511  \\
\enddata
\tablenotetext{a}{This backpropagation feature is available for the BA20 algorithm.)}
\end{deluxetable}

\begin{deluxetable}{lr}
\tabletypesize{\footnotesize}

\tablecolumns{2}
\tablewidth{0pc}
\tablecaption{Combination of \umlaut\  and BA20 algorithms: Fine-tuning of the P(\oiii)/P(\oii) ratio \tablenotemark{[a]}}
\tablehead{\colhead{$<\lambda_{\mathrm{obs}} [\emph{\AA}]>$} & \colhead{$\log$P$^{\mathrm{best}}_{\mathrm{[OIII]}/\mathrm{OII}}(\lambda_{\mathrm{obs}})$} }
\startdata
\label{tbl:Empirical_corr_OIII_OII}
8000  & 3.2  \\
9500  & 0.3  \\
13000 & 1.0  \\
14500 & 4.5  \\
15300 & 8.0  \\
17000 & 18.0 \\
\enddata
\tablenotetext{a}{This backpropagation feature is available for the BA20 algorithm.)}
\end{deluxetable}

\subsubsection{Accuracy and uncertainties}

The overall accuracy obtained by combining the supervised (BA20) and unsupervised approaches (\umlaut) is 84.4\%. This value is higher than that measured by using each of the two techniques separately. In Table~\ref{tbl:ACCU_TEST_2} we report the accuracy measured on the \ha, \oiii\ and \oii\ subsamples. Almost all the indicators improve when the two techniques are combined. In particular, both completeness and accuracy improves with respect to BA20 for all the three subsamples.

The performance of the algorithms varies significantly when considering the \oii\ sample. In particular, in BA20, the \oii\ sample is recovered with 50\% completeness and 26.7\% accuracy. Instead, using \umlaut, the completeness drops to 12.5\%, but the accuracy improves to 66.7\%. If we combine the two methods, we obtain a completeness of 66.7\% and an accuracy of 29.1\%.
These differences are mostly due to the fact that only few spectra of the WISP survey (24) are characterized by an \oii\ emission stronger than any other line. Consequently, the \oii\ sample is not particularly suitable for training an ML algorithm.

 \begin{deluxetable*}{cclllllll}
 \tabletypesize{\footnotesize}
 \tablecolumns{9}
 \tablewidth{0pc}
 \tablecaption{Accuracy obtained by combining \umlaut, our unsupervised algorithm, with the supervised algorithm of BA20.}

\tablehead{ species/ & $\lambda_{\mathrm{obs}}$ & N$_{\mathrm{TGS}}$\tablenotemark{[a]} & N$_{\mathrm{I}}$\tablenotemark{[b]} & N$_{\mathrm{CI}}$\tablenotemark{[c]} & N$_{\mathrm{WI}}$\tablenotemark{[d]} & Completeness  & Contamination & Accuracy (Purity) \\
        Transitions  & [\AA] & & & & & (N$_{\mathrm{CI}}$/N$_{\mathrm{TGS}}$) & (N$_{\mathrm{WI}}$/N$_{\mathrm{I}}$) & (N$_{\mathrm{CI}}$/N$_{\mathrm{I}}$) }
 \startdata
 \label{tbl:ACCU_TEST_2}
 \bf{H$\alpha$}  & \bf{6564.5}          & \bf{1293} & \bf{1325} & \bf{1155} & \bf{170} & \bf{89.3\%}  &  \bf{12.8\%} & \bf{87.2\%} \\
 \bf{\oiii} & \bf{4960.3 - 5008.2} & \bf{949}  & \bf{901}  & \bf{755}  & \bf{146} & \bf{79.6\%}  &  \bf{16.2\%} & \bf{83.8\%} \\
 \bf{\oii}\tablenotemark{[e]} & \bf{3727.1 - 3729.9}  & \bf{24}   & \bf{55}   & \bf{16}   & \bf{39} & \bf{66.7\%}  &  \bf{70.9\%} & \bf{29.1\%} \\
 \hline
 \bf{Overall accuracy:} & \multicolumn{8}{l}{\bf{84.4\%}\tablenotemark{[f]}} \\
 \enddata
\tablenotetext{a}{ Number of sources, in the test ``gold'' sample (TGS), showing this transition as the most prominent emission line.)}
\tablenotetext{b}{ Number of sources, in the test ``gold'' sample, for which the algorithm associates to the possibility of this transition to be the responsible for the most prominent emission line measured, the highest relative probability (Identified, I).}
\tablenotetext{c}{ Number of sources, in the test ``gold'' sample, for which the strongest emission due to this transition is correctly identified (CI).}
\tablenotetext{d}{ Number of sources, in the test ``gold'' sample, for which the strongest emission is mistaken for this transition (wrongly identified, WI).}
\tablenotetext{e}{ Due to the small amount of data included in the training and testing sets, the precision estimated should be considered an upper limit to the actual value.}
\tablenotetext{f}{ The ``gold'' sample includes a few lines that are not originally identified as \ha, \oiii\ or \oii. Because the algorithm can not be properly trained on these samples, we do not report the corresponding accuracies in this table.}
 \end{deluxetable*}

\begin{figure*}[!ht]
\centering
 \includegraphics[width=5.7cm]{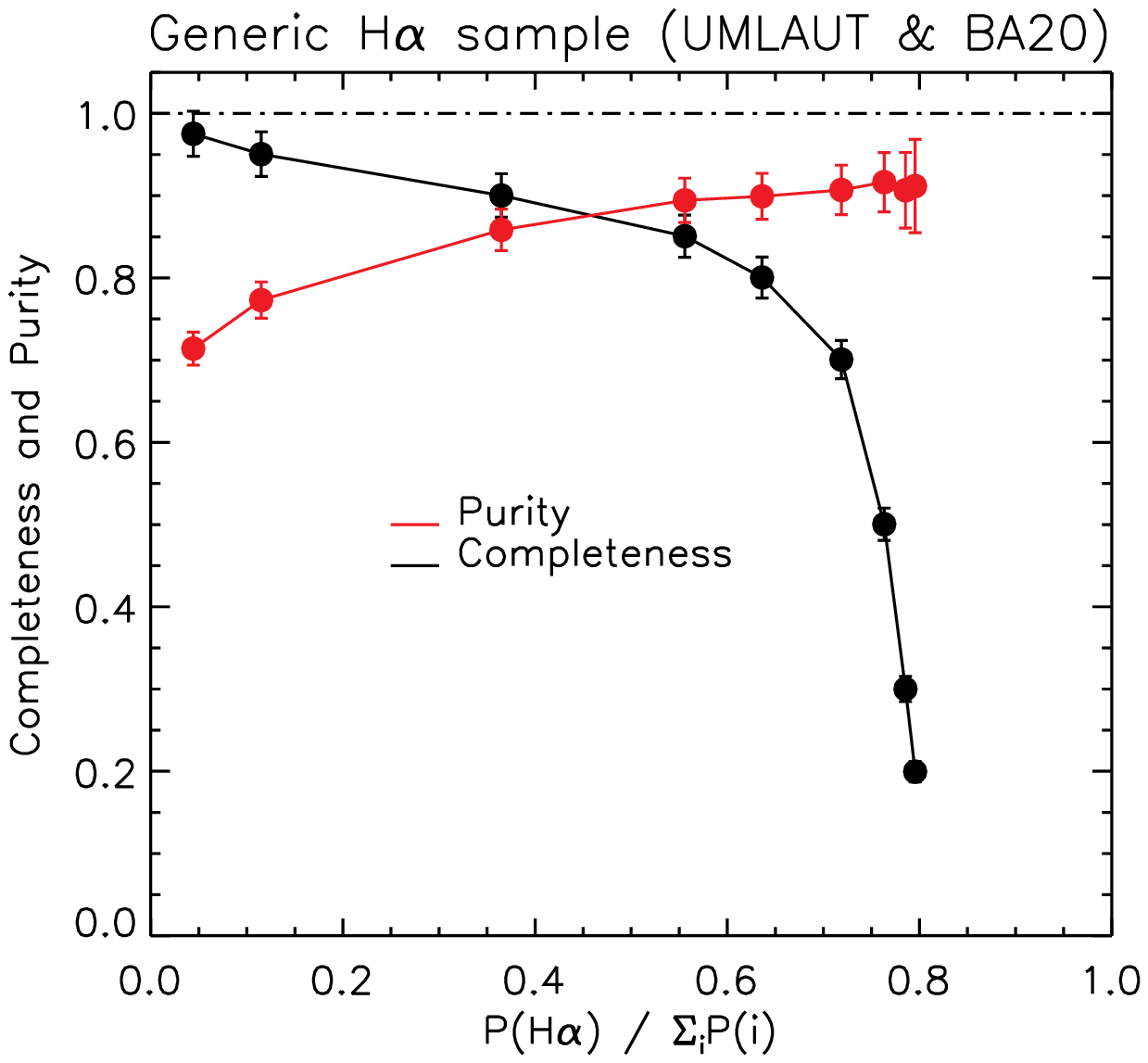}
 \includegraphics[width=5.7cm]{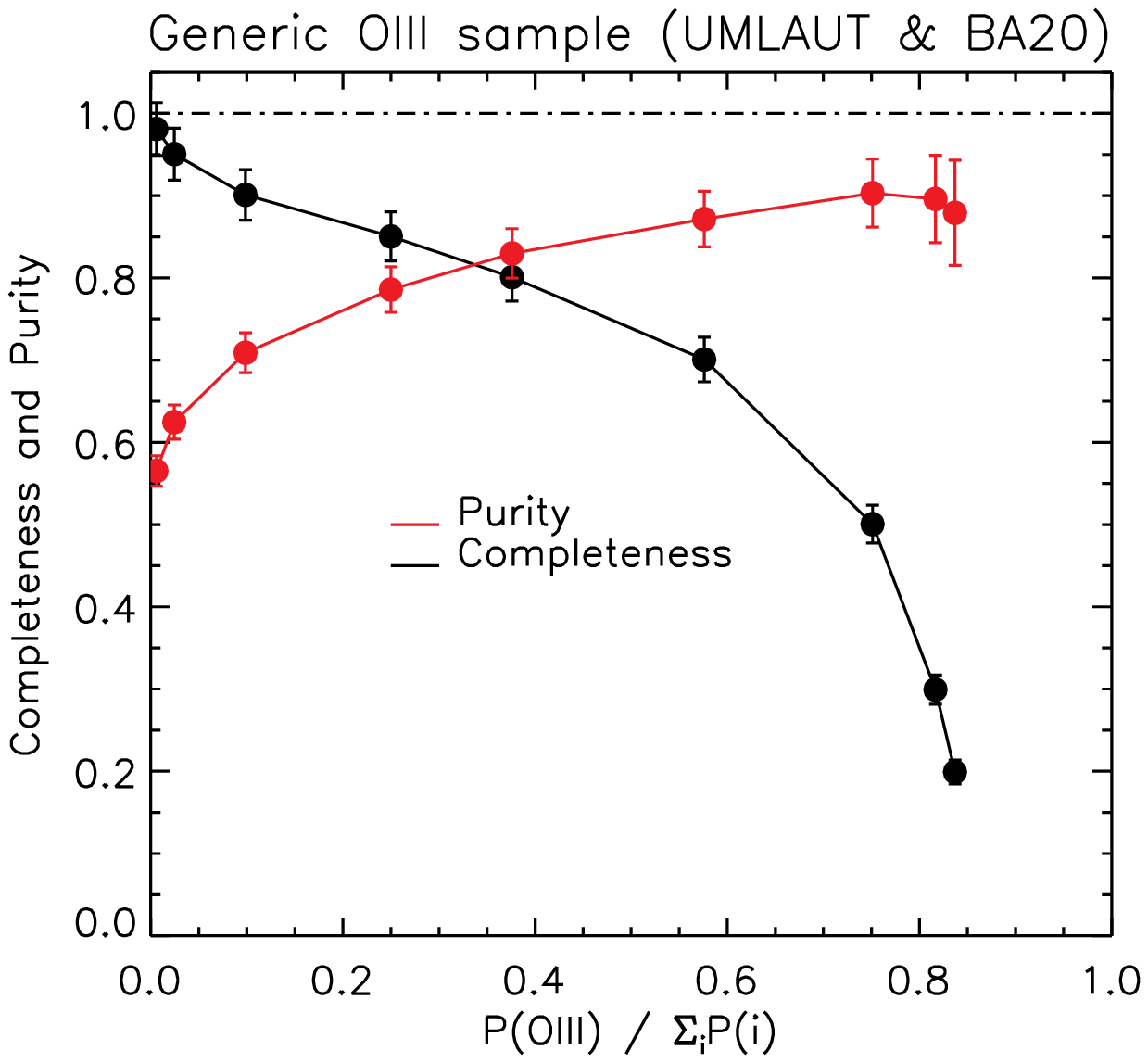}
 \includegraphics[width=5.7cm]{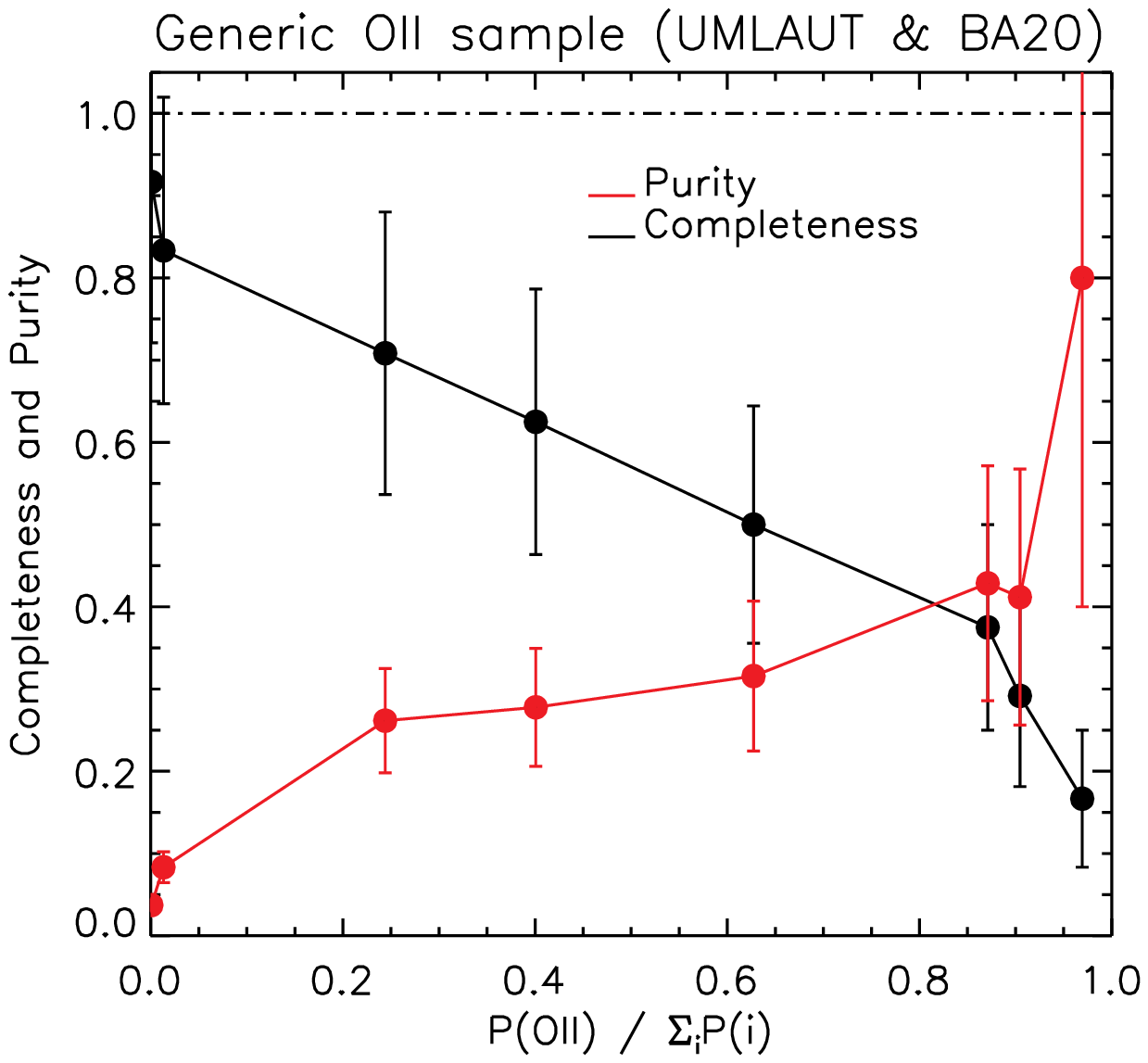}
 \caption{For every data point under analysis and for each of the classes considered, we compute one probability indicator: P(\ha), P(\oiii), and P(\oii). In these plots, the probability indicators are obtained from the combined PDF obtained from \umlaut\ and BA20. In our analysis, each line is associated with the label (\ha, \oiii, and \oii) characterized by the highest value of the corresponding probability indicator. This approach guarantees the highest overall accuracy, but it is not the only approach that can be adopted. In these plots we show how the probability indicators (computed from the outputs of \umlaut) can be fine-tuned to select more or less complete/pure samples of sources dominated by the \ha\ ({\bf left panel}), \oiii\ ({\bf central panel}), and \oii\ ({\bf right panel}) emission. In these examples, P(\ha), P(\oiii) and P(\oii) are normalized to the sum of the probabilities computed for all the species considered ($\sum_{i}$P(\emph{i})). The error bars correspond to Poissonian uncertainties.}
 \label{img:PROB_VS_PRECISION2}
 \end{figure*}

 The results reported in Table~\ref{tbl:ACCU_TEST_2} are obtained by classifying the lines in accordance with the highest probability indicator among P(\ha), P(\oiii) and P(\oii). However, Figure~\ref{img:PROB_VS_PRECISION2} shows how, by setting different thresholds on the same probability indicators, we can obtain samples characterized by different values of completeness and purity. Hence, as for the outputs of \umlaut\ only (Figure~\ref{img:PROB_VS_PRECISION1}), combining the outputs of \umlaut\ and BA20, we can still exploit the same possibility.

\newpage

\subsection{Identification of single lines in the WISP survey}
\label{SECT:Single_line_classification}
\begin{figure*}[!ht]
 \centering
 \includegraphics[width=8.9cm]{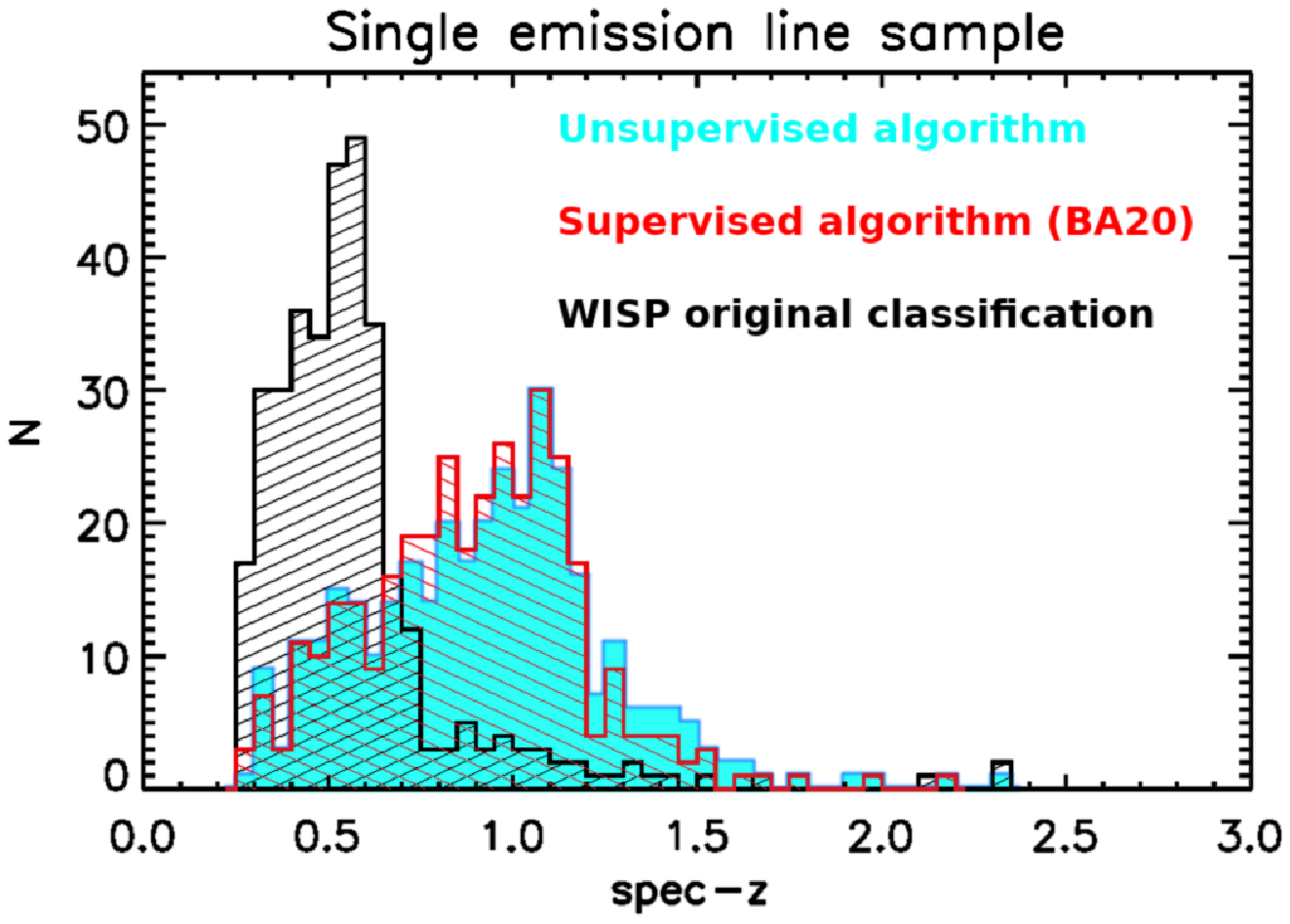}
 \includegraphics[width=8.9cm]{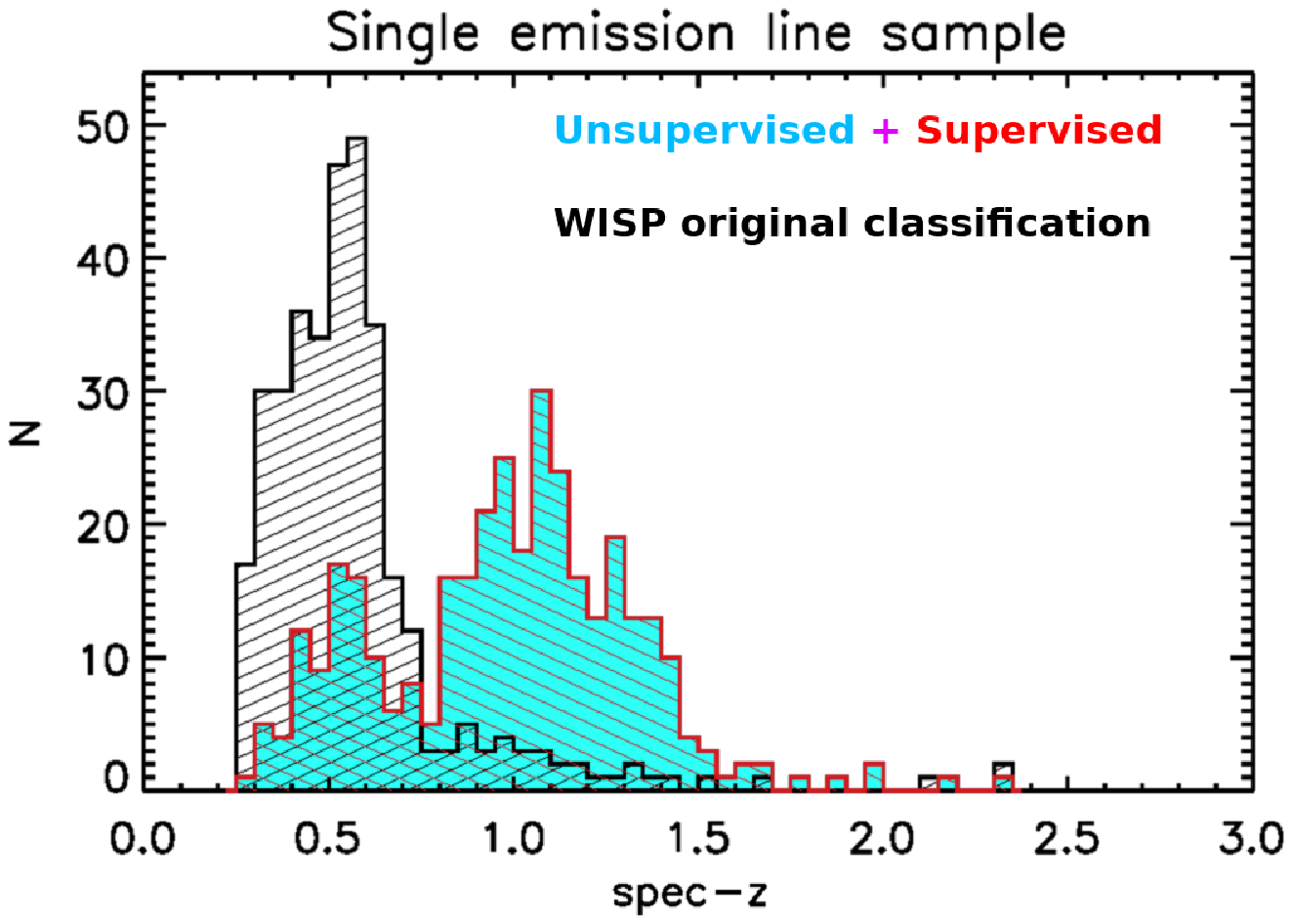}
 \caption{{\bf Left panel:} redshift distribution of the single-line sample following the original WISP classification (black histogram), considering the supervised classification of BA20 (red), and assuming the classification performed by \umlaut, our unsupervised algorithm (cyan). Our results substantially confirm the classification performed by the supervised algorithm of BA20. {\bf Right panel:} redshift distribution of the single-line sample obtained by combining our unsupervised approach with the supervised algorithm described of BA20.  }
 \label{img:z_DISTRIB}
 \end{figure*}

As in BA20, we used \umlaut\  to classify single emission lines detected in the WISP survey. In the left panel of Figure~\ref{img:z_DISTRIB}, we compare the redshift distribution obtained using our unsupervised algorithm with that resulting from the supervised classification of BA20. The distribution due to the original WISP classification is also shown in the background.

Almost all of the single emission lines are originally identified as \ha\ due to a default classification initially adopted by the WISP reviewers. 
As discussed in BA20, this default choice is probably well suited for bright low-$z$ sources, whereas it creates a strong bias when considering faint high-$z$ sources. 
In fact, they found that most of the single lines are actually \oiii\ lines, with 30\%-35\% of this sample due to spurious detections (usually classified as \oiii\ by the BA20 algorithm). As the redshift distributions of Figure~\ref{img:z_DISTRIB} show, \umlaut\  mostly confirms the classification of BA20.

  In the right panel of Figure~\ref{img:z_DISTRIB}, we show the redshift distribution of the single-lines sources obtained by combining our unsupervised algorithm with the unsupervised algorithm of BA20. In this case, a gap in the distribution clearly emerges at $z\sim$0.7. The same feature is also evident in the redshift distribution of the reference (gold) sample (see the left panel of Figure 20 in BA20) and it is due to the \ha\ line falling in the noisy region of the HST-G102 grism at these redshifts. The visibility of this feature further confirms the increased accuracy obtainable when combining supervised and unsupervised ML strategies.

\section{Discussion}
\label{SECT:DISCUSSION}

\subsection{The challenge of analyzing increasingly extensive datasets}

The WISP survey, which we used as a test bench for our unsupervised ML algorithm, represents one of the closest existing datasets to the future Euclid and Roman spectroscopic surveys. While we refer to BA20 and \cite{2020ApJ...897...98B} for an extensive discussion of this topic, here we just remind that these and other future surveys are expected to produce a tremendous amount of data, if compared with present-day and past surveys. The classification and analysis of such large datasets will no longer be performed by small groups of scientists: completely new strategies and methods will be required for this purpose.

In this context, two main paths can reasonably be followed. On the one hand, the solution could be offered by crowd-sourced science projects, such as those allowed by the zooniverse online interface\footnote{https://www.zooniverse.org} \citep[see e.g., ][]{2018RNAAS...2..120D}. On the other hand, automatic or semiautomatic approaches, and in particular ML strategies, can be used.

Using ML methods overcomes some of the problems associated with the citizen science strategy, such as those related to the reproducibility of the results, the lack of control on the experimental conditions, the availability of volunteers, etc. In any case, ML techniques can be combined with citizen science strategies: for example an ML algorithm can be trained using reference samples previously analyzed by citizen scientists \citep[see, e.g.,][]{2020MNRAS.491.1554W}.

\subsection{Advantages and limitations of our unsupervised algorithm}
\label{SECT:UMLAUT_PROS_CONS}

\umlaut\  represents an example of a general-purpose algorithm that can successfully be applied to large datasets of very different natures. In this paper, we have shown how it can reach accuracies similar to those obtainable using supervised ML strategies, but with the following relevant advantages:
\begin{itemize}
\item{the algorithm is immediately adaptable to very different contexts;}
\item{very few parameters need to be set by the user;}
\item{there is no need to know which of the input dimension is actually related to the output parameter and how;}
\item{the outputs can easily be integrated with those obtained using different algorithms;}
\end{itemize}

Compared to the algorithm presented in BA20, it is also worth mentioning how our unsupervised algorithm does not require any definition or use of internal functions. In BA20, a set of probability distribution functions independently relate each of the input parameters to the output ($z$). Hence, a final PDF is computed from the product of the many independent PDFs (see Figure 15 in BA20). This approach allows the user to extrapolate a lot of information from the combination of functions that are poorly informative when considered individually. On the other hand, considering one-dimensional functions has some drawbacks.

\begin{figure}[!ht]
 \centering
 \includegraphics[width=8.7cm]{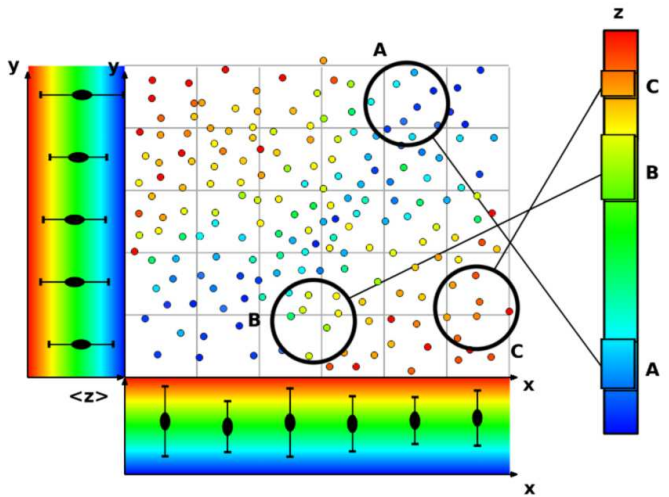}
 \caption{In this example, the $x$ and $y$ coordinates are not singularly related with $z$: both the $z(x)$ and the $z(y)$ functions (left and bottom plots, respectively) are flat. The $z$-PDF computed as the product of the two functions would be correspondingly flat (i.e. not informative with respect to the parameter $z$). On the contrary, given an ($x,y$) pair (any of the three black circles, corresponding to as many examples) it is possible to precisely estimate $z$ as the average $z$ measured on the closest $k$ elements ($k=7$, 8, and 9 in the three examples) in the $N$-dimensional space (here $N$=2)  }
 \label{img:ML_advantage1}
 \end{figure}

Figure~\ref{img:ML_advantage1} shows an example in which the combination of one-dimensional PDFs would fail in extrapolating useful information from the data distribution proposed. In this example, knowing the values of $x$ and $y$ would not help to determine $z$, as $x$ and $y$ are not individually related to $z$. \umlaut\  (the black circles in Figure~\ref{img:ML_advantage1} illustrate three different cases corresponding to as many ($x,y$) pairs) would more effectively extrapolate information from the same distribution, as it is based on the identification of the closest elements in the $N$-dimensional space.

  In Section~\ref{SECT:coord_weigh} and in Appendix~\ref{SECT:AppendixB} we describe the strategy used by \umlaut\ to identify and properly weight the input dimensions that are actually related to the output parameter.
 This method allows us to overcome some of the limitations affecting other techniques, such as the Principal Component Analysis \citep[PCA,][]{PAPER_PCA_PEARSON}. In fact, the PCA approach  assumes 
orthogonal native axes and Gaussian data distributions along them. Additionally, an underlying linear subspace is assumed \cite[][see in particular Figure 6 therein]{2014arXiv1404.2986S}. None of these assumptions is required by \umlaut.

As with many other ML algorithms, the accuracy of the outputs is strictly related to the characteristics of the training set. When the outputs of \umlaut\  are used to label objects belonging to different classes (as we do in our analysis for the \ha, \oiii, and \oii\ subsamples), 
 two main requirements are needed. First, the size of the smallest subsample used for training the algorithm should be large enough to allow for an acceptable accuracy on data of the same class. Second, the input data should natively encapsulate information on the class or subsample which any data point owns. In other words, \umlaut\  (as any other ML algorithm) can simply help \emph{extract} hidden information from a data set, but it does not \emph{create} information.

 In our accuracy tests (Table~\ref{tbl:ACCU_TEST_1}) we can see that both the completeness and the accuracy reached on the \oii\ sample are particularly low. This is due to a combination of the two factors just mentioned. In fact, only 24 spectra are characterized by an \oii\ line stronger than \ha\ or \oiii. Additionally, some of the redshift tracers considered (such as apparent magnitude and size) work pretty well only up to certain redshifts, making it difficult to correctly classify strong \oii\ emitters above $z\sim$1.5. This limitation is not due to how \umlaut\  processes the input information. Instead, it can mostly be attributed to the quality of the training set and in particular to the intrinsic amount of information that it encapsulates.

 In our tests, \umlaut\  is not trained on the identification of false detections. The WISP ``gold'' sample that we use to train the algorithm is made of real lines only. In these spectra, the real nature of the strongest lines and their identification are confirmed by the detection of additional lines (in any case, the algorithm is blind to their presence). Instead, the sample of WISP ``single lines'' classified by the algorithm (Section~\ref{SECT:Single_line_classification}) includes some false detections (BA20 quantify their amount to $\sim$30-35\%).
 
 In general, \umlaut\ should not be used to compute the unknown parameter of data belonging to categories it has not previously been trained on. This rule is valid unless the position of the data points belonging to the new category, in the ($N+1$)-dimensional parameter space, overlaps with the position of some of the data used for the training. Therefore, if \umlaut\ is trained using only real lines and then it is applied to false detections, then the algorithm will identify the real lines of the training set that more closely resemble each of the false detections under analysis. Then, it will try to compute the output parameter (i.e., the redshift) as if the false detection under analysis were a real line. On the contrary, \umlaut\ can be trained on simulated data and then applied to real analysis data (and vice versa), as these two different categories overlap in the ($N+1$)-dimensional parameter space.

 It is important to note that \umlaut\ (at least in its current version) has a limited capability to extrapolate the proximity relations learned from the training set, outside the multidimensional parameter space defined by the training set itself. This is due to the design of \umlaut, which estimates the unknown parameter directly from the training data that more closely resemble the analysis data point. This approach is followed regardless of whether the data point under analysis is located inside or outside the boundaries of the multidimensional space defined by the training dataset.

 For example, if we were trying to estimate the weight of an adult elephant by exploiting the physical properties ($N$ parameters) of a training set made of young elephants only, then \umlaut\  would return the average weight of the heaviest young elephants in the training sample, (i.e., the weight of the young elephants that most resemble the adult elephant under analysis). In the example described, the estimation would clearly be inaccurate, but it must be understood that it would also be the most accurate possible, \emph{given the training set used}.

 The ``fit'' option (see Section~\ref{SECT:K_CLOSE}) allows \umlaut\  to overcome some of the problems related to the analysis data points located at (or just outside) the edges of the multidimensional parameter space defined by the training set. However, the same feature should not be expected to provide correct/accurate results when the range of values covered by the training dataset along the unknown $(N+1)th$ dimension does not encompass the expected value for the analysis data point. Additionally, \umlaut\  is expected to provide less accurate results when the properties of the analysis dataset (i.e., the ranges of values covered by the training set along the $N$ dimensions) substantially differ from those characterizing the training set.

 The limitation just described affects the algorithm of BA20 less heavily as, in that case, every function can be extrapolated (up to a reasonable point) outside the range of values covered by the training set. This advantage of the BA20 algorithm comes at the expense of the advantages of \umlaut, listed at the beginning of this section. With this final remark, we stress the potential advantages provided by the combination of complementary approaches such as the two techniques that we compared in this paper.

\subsection{\umlaut\ in the context of the Euclid and Roman surveys}
\label{SECT::UMLAUT_EUCLID_ROMAN}

 In Section~\ref{SECT:UMLAUT_PROS_CONS}, we outlined some of the advantages and limitations characterizing our algorithm. Following that discussion, in this section, we propose some examples of the use of \umlaut. The methods described here could be effectively adopted to classify single spectral lines, especially in the context of the Eulcid and/or Roman surveys.

 The highest level of accuracy allowed by \umlaut\ can be obtained if a subsample of single-line spectra, taken from the same survey, can be used as a training set. Of course, these spectral lines should already be independently identified, and this is possible only if additional spectroscopic and/or photometric observations are available for this subsample. Samples of the type just described are already available for some cosmological fields, while others will follow after the realization of the Euclid and Roman surveys themselves.

 Alternatively, in accordance with the analyses performed in this paper, \umlaut\ can be trained on spectra characterized by multiple lines detected (in this case, the line identification is almost univocal and more easily automatable) and then directly applied to an analysis sample made of single-line spectra. Also in this case, both the training and analysis sets would be sampled from the same survey, keeping the accuracy at high levels.

 Finally, given the similarities between WISP and the future Euclid and Roman surveys, the same data presented in this paper could be used as a training set. In this case, the training and analysis sets would not be sampled from the same surveys, probably generating less accurate results. Therefore, while this approach is still feasible, the first two methods are preferable.

\section{Summary and conclusions}
\label{SECT:SUMMARY_AND_CONL}

In this paper, we described \umlaut, a new freely available software (https://github.com/Ibaronch/UMLAUT). \umlaut\ is an unsupervised ML algorithm primarily designed for (but not limited to) regression purposes.
We tested the algorithm  on a ``gold'' sample of spectra collected in the context of the WISP survey, one of the most important proxies of the future Euclid and Roman missions.

All the lines considered in our test represent the brightest feature detected in each spectrum. Their secure identification is made possible by the simultaneous detection of additional lines to which the algorithm is always blind.
We use \umlaut\  to estimate the redshift of the emitting sources. Hence, knowing the observed wavelength of the strongest emission line, its identification is obtained using the output $z$-PDF.
We obtain an overall accuracy of 83.2\%, which is comparable with that measured by BA20 (82.6\%) by applying a supervised algorithm to the same sample. The accuracy increases to 84.4\% when the two techniques are combined.

The configuration of our test  makes it possible to use the algorithm for the identification of single spectral lines when only the strongest one is detected (the weaker lines being below the detection threshold). When applied to the WISP single-line sample, \umlaut\  substantially confirms the classification previously reported in BA20.

\umlaut\  does not require internal functions to be set and, up to a certain limit, there is no need to know which of the input dimensions are related to the unknown output parameter(s). The algorithm automatically organizes the inputs and there is no need to rescale or modify their distributions.
Thanks to its high flexibility, \umlaut\  is suitable for use in a very wide range of different domains.
As an example of this flexibility, the first author participated in the linguistic research presented in \cite{10.3389/fpsyg.2019.01528}.
 In that study, we prepared a dataset that has already been analyzed by \umlaut\  with the goal of determining the geographical origin of contemporary and past speakers of a German heritage dialect (I. Baronchelli \& F. Cognola 2020, in preparation)\footnote{Some of the results of this work can already be found at https://zenodo.org/record/3484665\#.X9dpmcKnfMY}.

 The current version of \umlaut\ (\umlaut\ 1.0) is written in IDL (Interactive Data Language) and it is designed to run sequentially (one analysis data point after another) on normal CPUs (Central processing units). We are currently planning to rewrite some portions of \umlaut\ to allow users to run future versions of the software in parallel threads on GPUs (graphic processing units).


\appendix
\twocolumngrid 


\section{The apparently overlapping training and testing sets: alternative accuracy tests}
\label{SECT:AppendixA}

When testing ML algorithms, an original reference sample is usually divided into two independent samples, one of which is used to train the algorithm and the other to test it. In fact, if the algorithm is allowed to generate an arbitrarily complex set of internal rules that fully describe all the characteristics of the data points considered, any training dataset can be recovered with $\sim$100\% accuracy. Therefore, testing the algorithm on the same set used to train it generates the so-called ``overfitting'' problem (i.e. an overestimation of the supposed accuracy of the algorithm).

As mentioned in Sections~\ref{SEC:cal_test_samples} and ~\ref{SECT:WISP_TESTS},
\umlaut\  is tested on the same data points included in the training sample. 
However, as we will show, this practice does not generate overfitting problems because by following the leave-one-out validation strategy \citep[e.g.,][]{10.2307/1267500,doi:10.1111/j.2517-6161.1974.tb00994.x,hastie01statisticallearning}, the overlap between the two samples is only apparent.

ML algorithms can effectively be tested by dividing the original training set into two subsamples. Hence, the accuracy can be measured by training the algorithm on one of the two sets and testing it on the other.
The accuracy of the algorithm can be estimated even more precisely by randomly distributing the data of the original sample between the two subsamples and repeating the same experiment many times, averaging the results. While such a strategy should prevent ``overfitting'' problems, it must be noted that the resulting estimation always represents lower limits to the actual accuracy of the algorithm, as the algorithm itself is always trained on only a fraction ($\sim 0.5$) of the available data.

Following the `leave-one-out cross-validation'' strategy, \umlaut\  retraines itself every time it is tested on a new data point. When testing a given data point, the same datum is not used to train the algorithm. From this perspective, the actual testing set (i.e. the data point tested) and training set (i.e. the reference sample from which the data point under analysis has been removed) are always fully independent. 
The advantage of the `leave-one-out cross-validation'' strategy is that the entire reference sample (excluding the tested data point) can be used to train the algorithm, allowing for a more precise estimation of the maximum accuracy reachable by the algorithm.

In order to convince the skeptical reader, we performed 
some comparative runs  determining the accuracy of \umlaut\  by dividing the original ``gold'' sample into two fully independent subsamples, one of which is used to train and the other to test the algorithm). 
In particular, we performed 12 independent tests. Before each of these tests we randomly mixed the data points, creating a new pair of independent sets. At the end of the process, we obtained an average accuracy of 81.70\%$\pm$0.07\% (with 0.83\% dispersion). This value is just a little bit lower than using the ``leave-one-out validation'' (83.2\%, see Table~\ref{tbl:ACCU_TEST_1}). However, as said above, the residual discrepancy observed is expected, as training the algorithm on only half of the available ``gold'' sample lowers the resulting accuracy. This circumstance is proven by some additional tests that we performed as described here below.

If we follow the ``leave-one-out validation'' strategy but considering only half of the complete ``gold'' sample, we measure an average accuracy of  81.93$\pm$0.06 (with 0.76\% dispersion). Again, we obtain this value as the average of 12 different tests where the data points in the training/test set are randomly selected from the original sample.
As expected, this level of accuracy is consistent with that obtained by dividing the  ``gold'' sample into two independent samples, as in the 12 previous tests. At the same time, the accuracy is lower, if compared to that obtained following the same ``leave-one-out validation'' strategy but considering the entire ``gold'' sample).
In other words, when the algorithm is trained using the same amount of data, the accuracy does not sensibly change if we follow the ``leave-one-out validation'' strategy, or if we divide the ``gold'' sample into two numerically similar and independent training and testing sets.

\section{Simulation on coordinates weighting with uncorrelated inputs}
\label{SECT:AppendixB}

In order to test the ability of \umlaut\  to properly weight the input parameters, we performed a series of additional tests on the WISP ``gold'' sample (Section \ref{SEC:cal_test_samples}). 

As for the standard runs discussed in the paper, in all these tests we included the following input parameters ($N$ input dimensions):
\begin{itemize}
\item{the apparent magnitude in the F110W ($J$) band;}
\item{the apparent magnitude in the F140W ($H$) band;}
\item{the apparent magnitude in the F160W ($H$) band;}
\item{the apparent size;}
\item{the F110W-F140W color index;}
\item{the F110W-F160W color index;}
\item{the equivalent width (EW) of the spectral line;}
\item{the observed wavelength $\lambda$ of the emission line.}
\end{itemize}
Because these inputs are not always available at the same time, the actual value of $N$ ranges from $N=4$ to $N=6$. The correlation existing between each of these quantities and the redshift of the emitting source is shown in BA20.

\begin{figure*}[!ht]
 \centering
 \includegraphics[width=8.9cm]{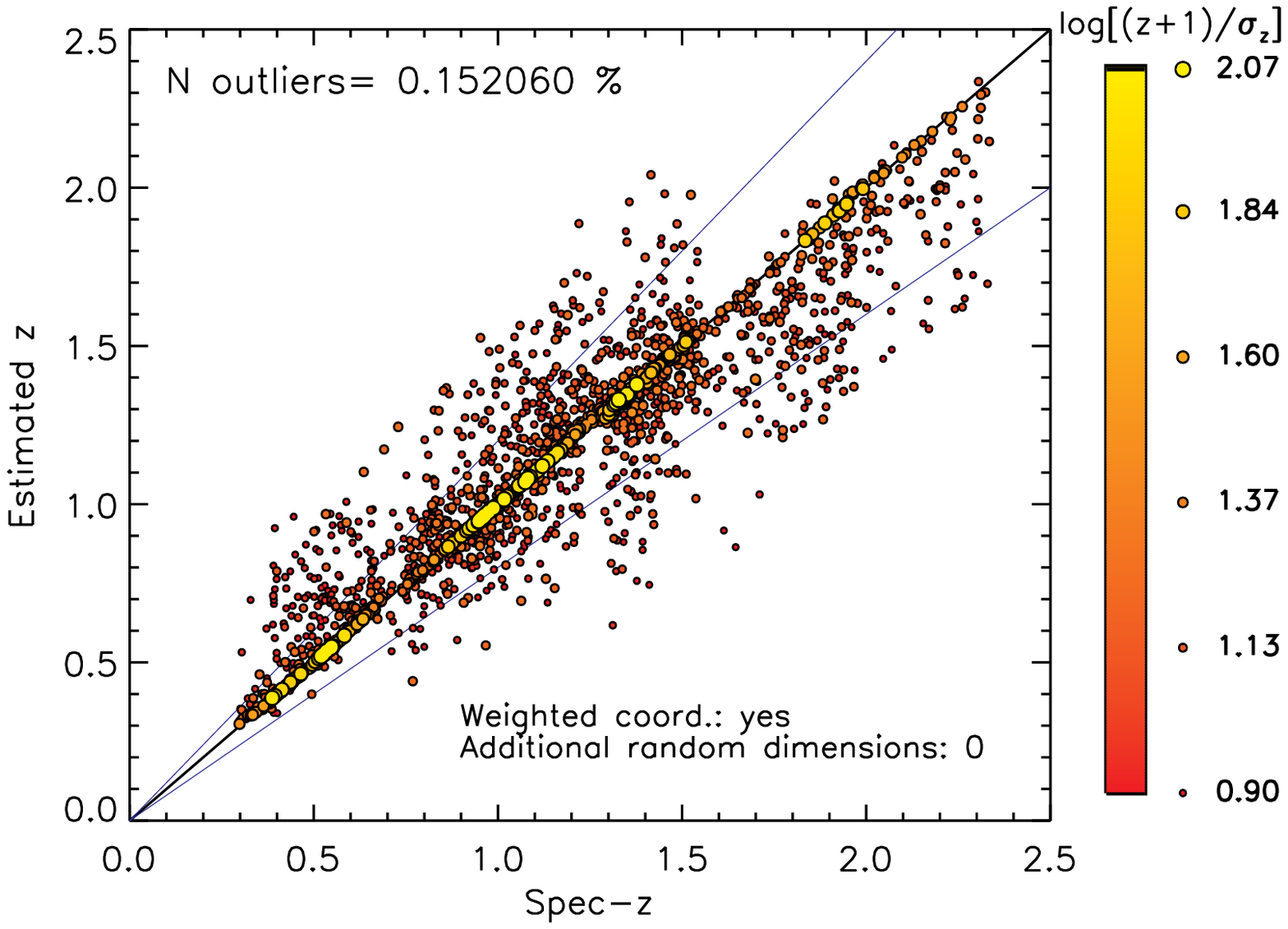}
 \includegraphics[width=8.9cm]{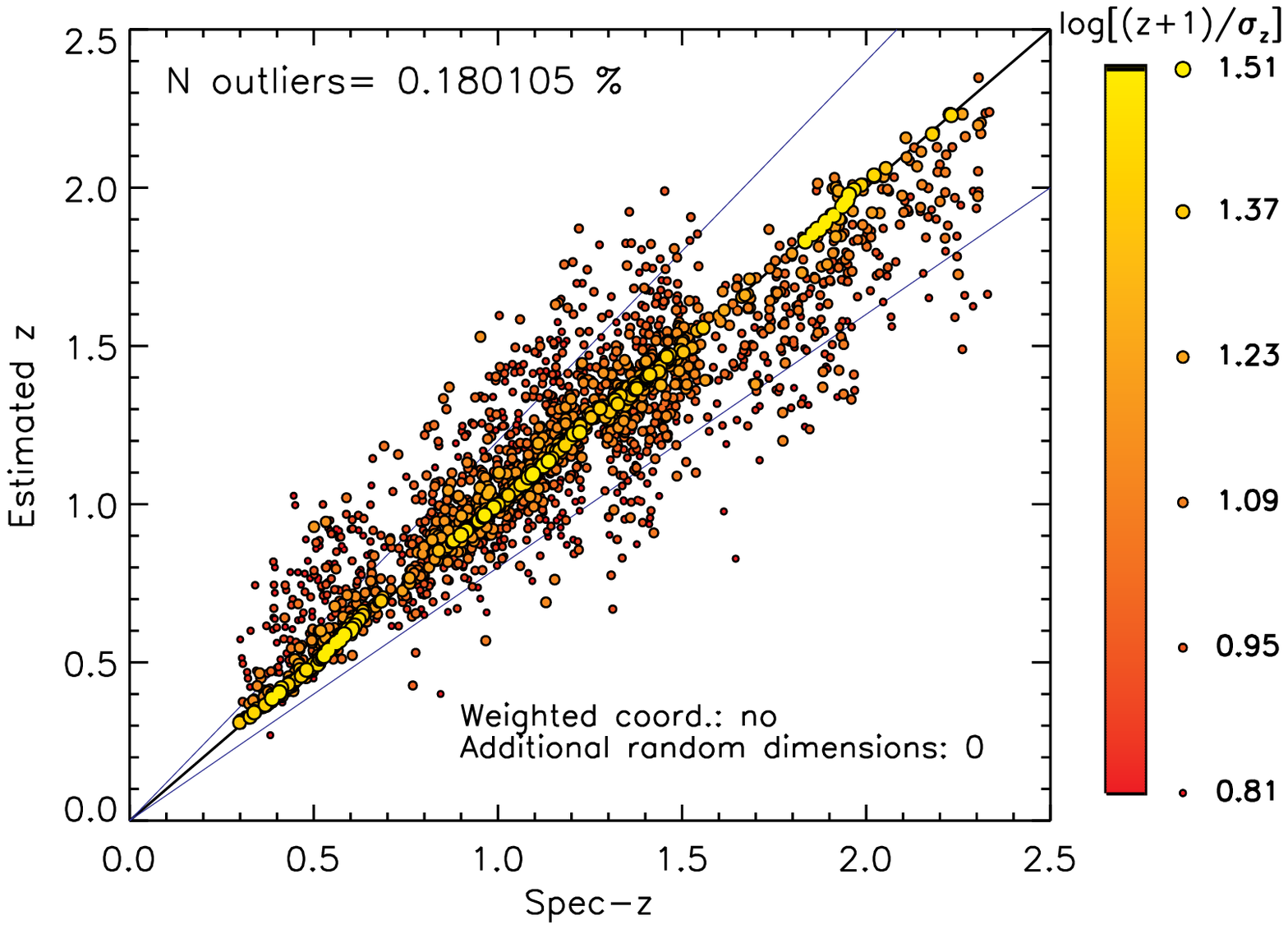}
 \includegraphics[width=8.9cm]{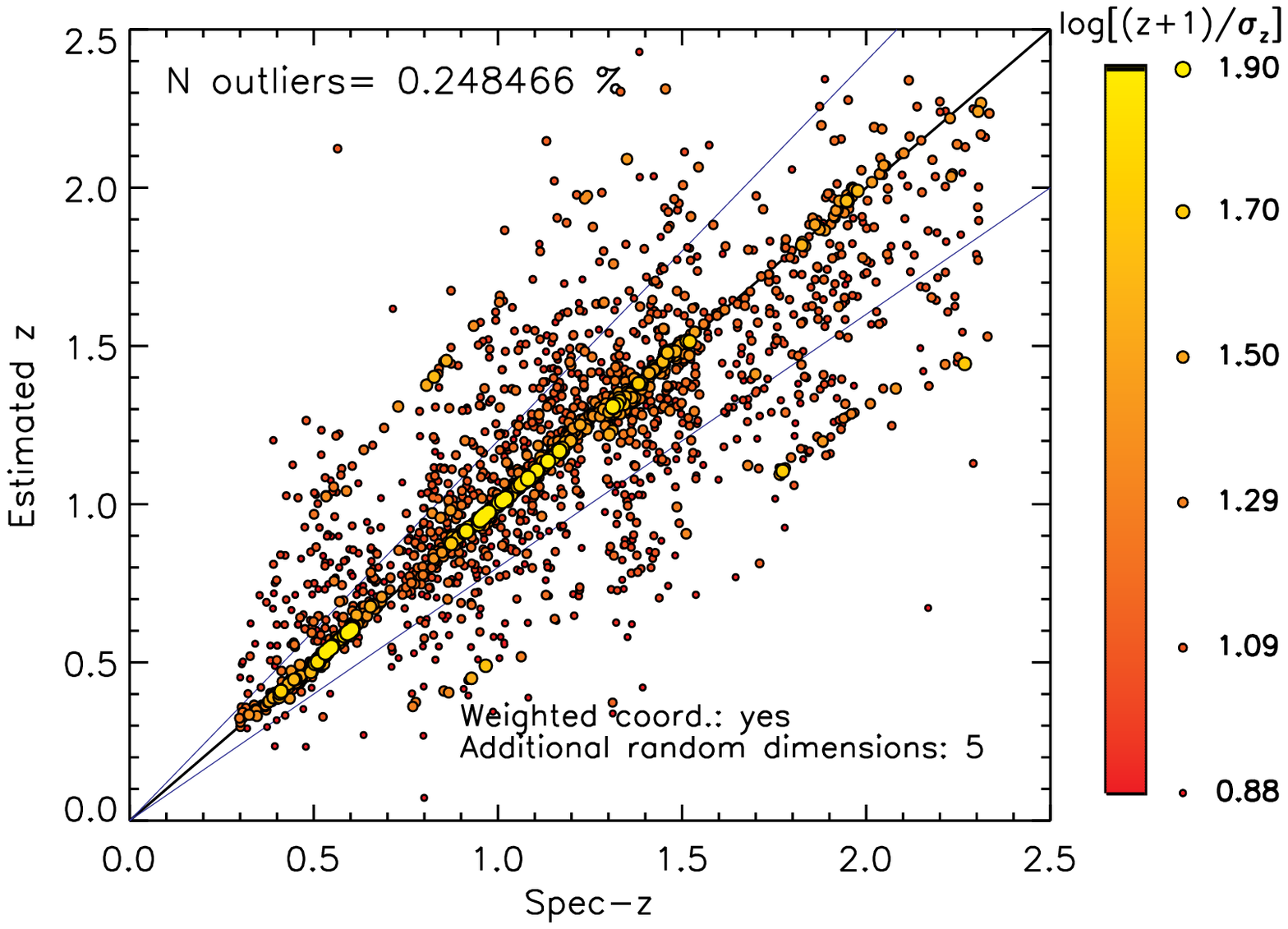}
 \includegraphics[width=8.9cm]{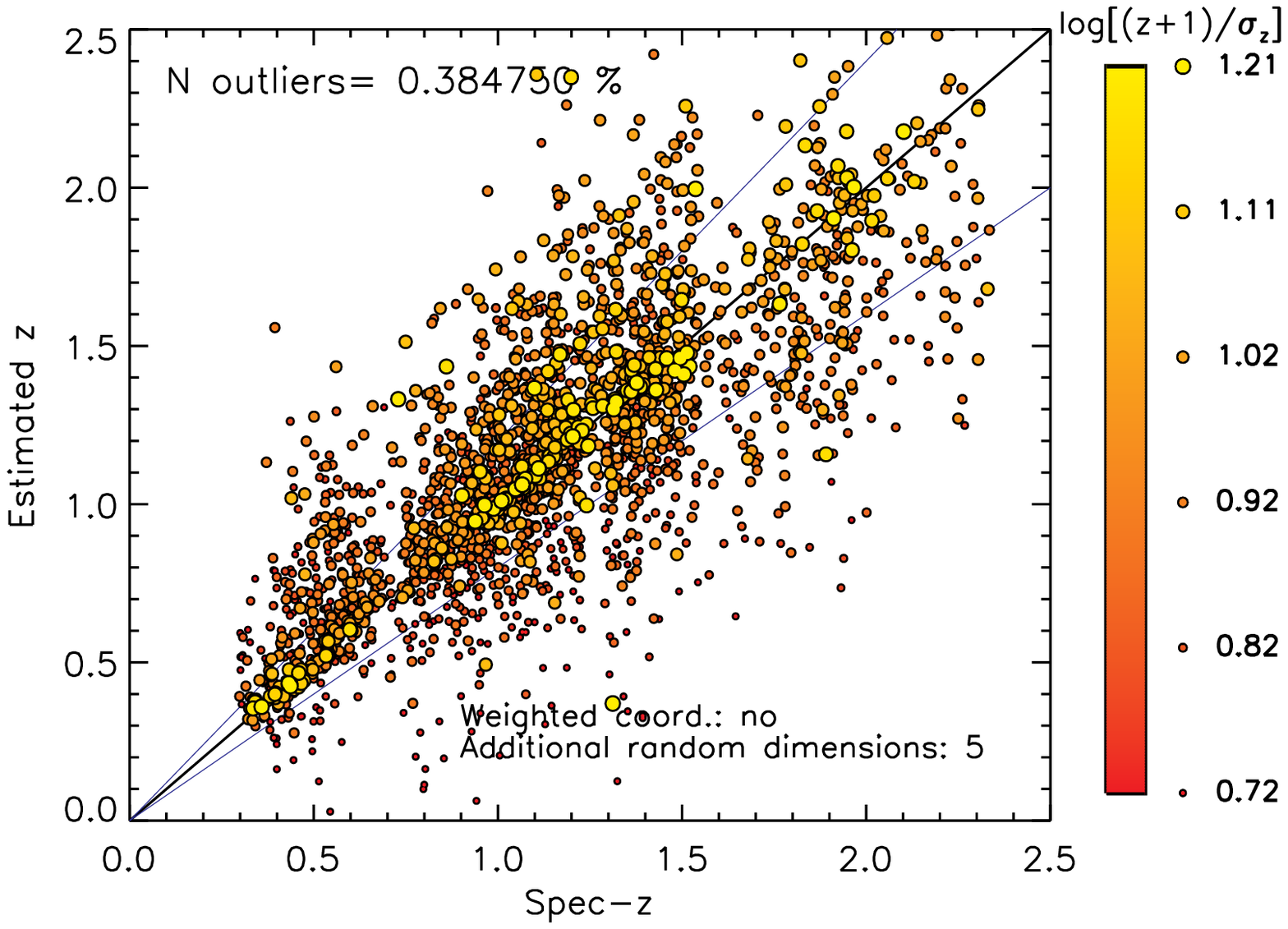}
 \includegraphics[width=8.9cm]{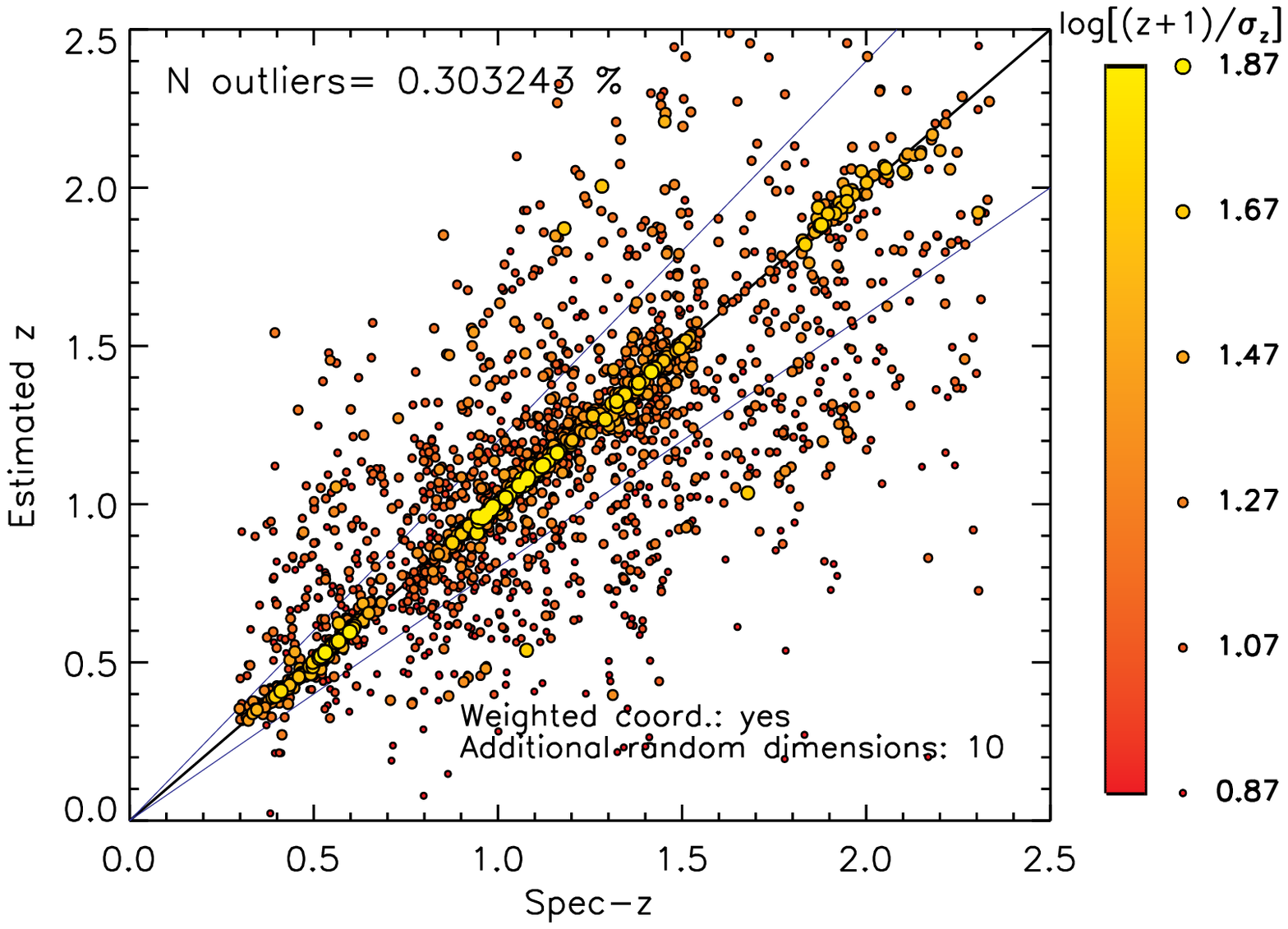}
 \includegraphics[width=8.9cm]{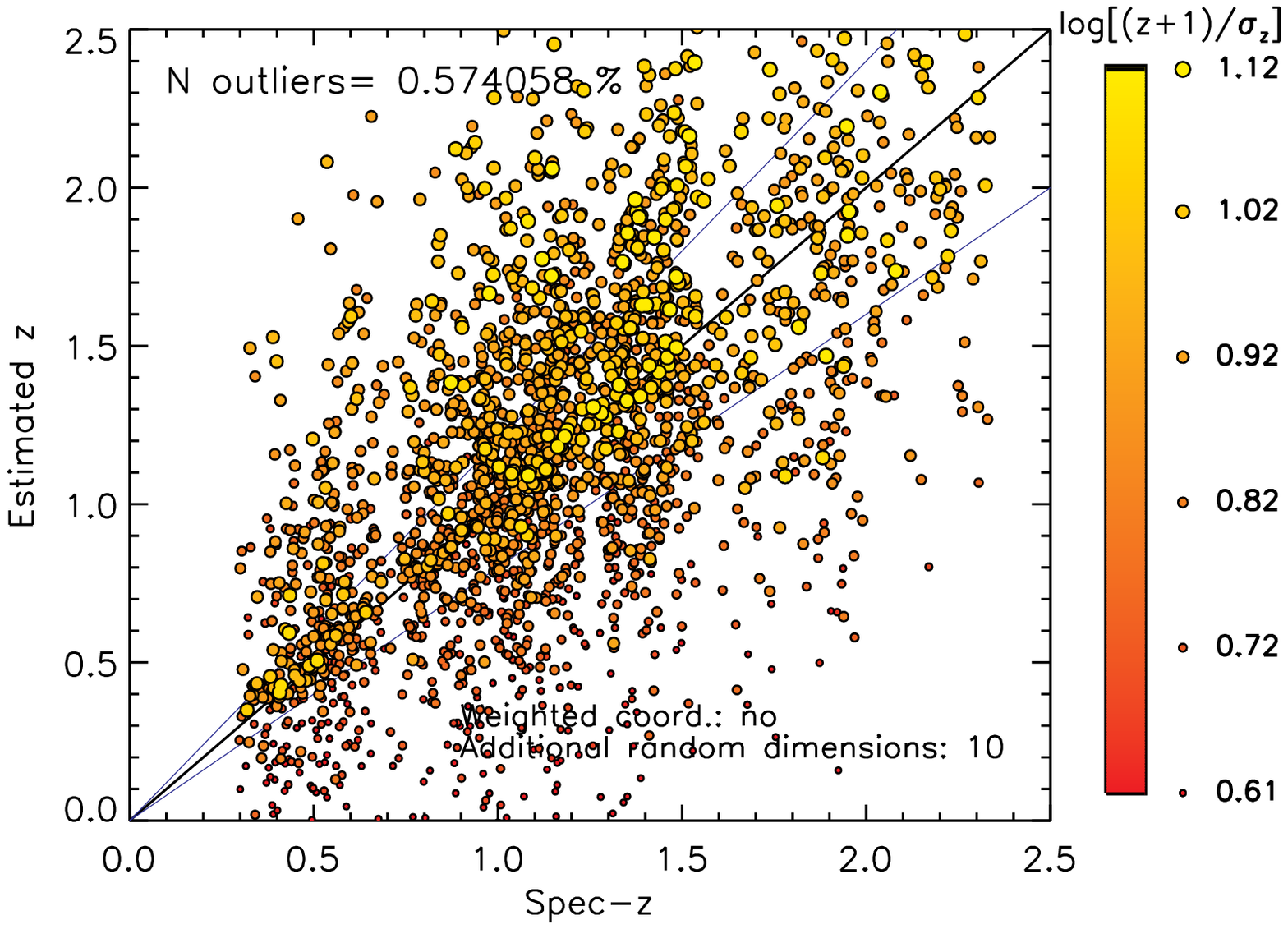}
 \caption{Redshift estimated by our unsupervised algorithm vs. actual spectroscopic redshift. The output redshift is obtained using the multiple linear regression option (see Section~\ref{SEC:OUTPUTS}). In the {\bf left panels}, we show the results obtained by allowing the algorithm to properly weight the rank-based input coordinates. In the {\bf right panels} the weighting option is not active and the number of closest data points considered ($K$) is fixed. The difference between the two configurations is not dramatic when all the input dimensions are correlated with the output parameter ({\bf upper panels}). On the other hand, the advantage of using weighted rank-based coordinates is evident when half ({\bf middle panels}) or even 2/3 ({\bf lower panels}) of the input dimensions are not related to the output parameter. In these examples, we considered additional input dimensions where the values assumed by the data are randomly extracted from normalized Gaussian distributions. The uncertainty associated with the output parameter, estimated by the algorithm, is shown using a color/dimension scale, with bigger yellow circles corresponding to smaller uncertainties and smaller red dots corresponding to higher uncertainties. This scale is not preserved among  the different plots.}
 \label{img:test_weight_coord1}
 \end{figure*}

In an initial test, we left \umlaut\  free to weight the input dimensions and to select the number of closest elements to use (15$<K<$25). Basically, we set the algorithm as already described in Section~\ref{SECT:UNSUP_ALG_ON_WISP}, with the only difference that, in this case, we do not compute an output $z$-PDF. Instead, we just consider the output value of $z$ estimated through a multiple linear regression (see Section~\ref{SEC:OUTPUTS}). The results of this initial test are shown in the upper-left panel of Figure~\ref{img:test_weight_coord1}.

In our second test, we did not allow \umlaut\  to weight the rank variables of the parameter space. Additionally, the number of closest elements ($K$) is kept fixed to 20. The result of this test is shown in the upper right panel of Figure~\ref{img:test_weight_coord1}.

It is possible to visually appreciate that the precision of the output is not dramatically compromised when the weighting option is not active.
In place of measuring the actual dispersion of the data, let us define the outliers as sources characterized by:
\begin{eqnarray}
z_{\mathrm{est}}-z_{0}>z_{0}*0.2
\end{eqnarray}
(with $z_{0}$ and $z_{\mathrm{est}}$ corresponding to the actual and estimated redshift respectively). This definition roughly defines the threshold above which the line considered would be wrongly classified and better represents the accuracy of the outputs for non-Gaussian distributions.
Following this definition, the coordinate weighting reduces the fraction of catastrophic redshift estimations by about 3\%, from 18.0\% to 15.2\%. The relatively limited extent of the improvement is expected, as all the inputs considered are known to be related to the output parameter (redshift), as shown in BA20. 

Instead, the actual advantage of weighting the input coordinates becomes relevant if we include additional input dimensions that are not related to the output parameter. This situation represents well all those cases when we do not know which of the input dimensions is actually related to the output parameter and how.

In the middle panels of 
Figure~\ref{img:test_weight_coord1} we show the results of the tests described above when five additional dimensions for each of the reference data points are included. The value assumed by any of the input data point along each of these additional dimensions is randomly selected from a preset Gaussian distribution \footnote{The actual value assumed by each datum is not relevant, as the input dimensions are converted into rank variables (see Section~\ref{SECT:ORDINALIZATION}). What matters here is the reciprocal position of the data inside the distribution itself.}. This configuration corresponds to having half of the input dimensions not related to the output parameter.
In this case, when the coordinates are weighted (middle left panel), the algorithm is able to limit the effects of the added random dimensions. On the contrary, the precision quickly worsens if the coordinates are not weighted (middle right panel).

The difference between the two configurations dramatically increases when 10 uncorrelated additional dimensions ($\sim 2/3$ of the input dimensions) are considered (bottom panels of 
~\ref{img:test_weight_coord1}).
In this case, while the accuracy is still acceptable when weighting the input dimensions (left panel; $\sim$70\% correct identifications), the unweighted configuration brings unacceptably low accuracy (right panel, $\sim$43\% correct identifications)

These tests demonstrate the efficiency of \umlaut\  in weighting the input dimensions depending on their degree of correlation with the output parameter. It is important to notice that this degree of correlation is not measured in a two-dimensional input-output space, such as the $z=f(x)$ space, but in the $N$-dimensional parameter space. This characteristic allows \umlaut\ to obtain useful results also in particular cases such as those shown in Figure~\ref{img:ML_advantage1}, where the two-dimensional correlation would fail in weighting the input parameters.

\goodbreak
\bibliographystyle{apj}
\bibliography{biblio4}{}

\end{CJK*} 
\end{document}